\documentclass[%
 aip,
 floatfix,
 amsmath,amssymb,
 reprint,%
]{revtex4-1}

\usepackage{graphicx}

\usepackage{dcolumn}
\usepackage{bm}

\usepackage[utf8]{inputenc}
\usepackage[T1]{fontenc}
\usepackage{mathptmx}
\usepackage{etoolbox}
\usepackage{amsmath}

\newcommand{\id}{\mathbf{1}}

\newcommand{\rd}{\mathrm{d}}

\newcommand{\be}{\begin{equation}}
\newcommand{\ee}{\end{equation}}
\newcommand{\bes}{\begin{eqnarray}}
\newcommand{\ees}{\end{eqnarray}}
\newcommand{\ket}[1]{{\left|  #1 \right\rangle}}

\newcommand{\bra}[1]{{\left\langle  #1 \right|}}

\newcommand{\hide}[1]{{}}

\DeclareMathOperator{\arccot}{arccot}

\makeatletter
\def\@email#1#2{%
 \endgroup
 \patchcmd{\titleblock@produce}
  {\frontmatter@RRAPformat}
  {\frontmatter@RRAPformat{\produce@RRAP{*#1\href{mailto:#2}{#2}}}\frontmatter@RRAPformat}
  {}{}
}%
\makeatother

\begin{document}

\preprint{AIP/123-QED}

\title{Machine Learning Approach to the Floquet--Lindbladian Problem
}

\author{V.~Volokitin}
 \email{volokitin@itmm.unn.ru}
 \affiliation{Department of Mathematical Software and Supercomputing Technologies, Lobachevsky University, Nizhny Novgorod 603950, Russia}

\author{I.~Meyerov}
 \affiliation{Department of Mathematical Software and Supercomputing Technologies, Lobachevsky University, Nizhny Novgorod 603950, Russia}

\author{S.~Denisov}
 \affiliation{Department of Computer Science, Oslo Metropolitan University, Oslo N-0130 Oslo, Norway}
\affiliation{NordSTAR – Nordic Center for Sustainable and Trustworthy AI Research, Oslo N-0166, Norway}
\affiliation{Department of Applied Mathematics, Lobachevsky University, Nizhny Novgorod 603950, Russia}

\date{\today}

\begin{abstract}
Similar to its classical version,
quantum Markovian evolution can be either time-discrete or time-continuous. 
Discrete quantum Markovian evolution is usually modeled with completely-positive trace-preserving maps while time-continuous evolution is often specified with superoperators referred to as "Lindbladians". Here we address the following question: Being given a quantum map, can we find a  Lindbladian which generates an evolution identical -- when monitored at discrete instances of time -- to the one induced by the map? It was demonstrated that the problem of getting the answer to this question can be reduced to an NP-complete (in the dimension $N$ of the Hilbert space  the evolution takes place in) problem. We approach this question  from a different perspective by considering a variety of Machine Learning (ML) methods and trying to estimate their potential ability to give the correct answer. Complimentary, we use the performance of different ML methods as a tool to check the hypothesis that the answer to the question is encoded in spectral properties of the so-called Choi matrix, which can be constructed from the given quantum map. As a test bed, we use two single-qubit models for which the answer can be obtained by using the reduction procedure. The outcome of our experiment is that, for a given map, the property of being generated by a time-independent Lindbladian is encoded both in the eigenvalues and the eigenstates of the corresponding  Choi matrix.
\end{abstract}

\maketitle

\begin{quotation}
The question posed in the abstract is a quantum version of the "embedding problem" formulated by Elfving in 1937 \cite{elfing} for classical Markov processes: Given map $\mathcal{P}$ can we find  generator $\mathcal{L}$ such that $\mathcal{P} = \mathrm{exp}(\mathcal{L}T)$, where $T$ is the given time interval? Answer "yes" would mean that the original time-discrete evolution can be obtained from the constructed continuous-time evolution  by monitoring the latter at the time instances $t=T,2T,...$. It may also be  that the answer is "no" and the time-discrete evolution cannot be obtained as a stroboscopic sample of \emph{any} time-continuous Markovian evolution.

The  problem of finding the answer for a given completely positive trace-preserving (CPTP) map  $\mathcal{P}$ was called "[quantum] Markovianity problem"~\cite{X1}. An algorithm to obtain the answer, based on a reduction of the original problem  to a particular problem of integer semidefinite programming, has also been proposed~\cite{X1}. In its turn, this particular problem was shown~\cite{X2} to be reducible to the well-known NP-complete $\mathrm{1{\text -}IN{\text -}3{\text -}SAT}$ problem~\cite{np}. There is no surprise that the algorithm has an exponential complexity with respect to the  problem dimension $N$. A recent attempt to implement the algorithm demonstrated that the practical horizon is limited by $N = 8$ (e.g., by models consisting of no more than three spins/qubits)~\cite{unpub}.

Recently, the Markovianity problem has gained particular attention in the context  of open quantum evolution governed by time-dependent Lindbladians. While
the stroboscopic version of coherent quantum evolution, determined by a time-periodic Hamiltonian, can always be obtained with an effective  time-independent Floquet Hamiltonian~\cite{Holthaus2015,Bukov2015}, it is no longer so in the case of open quantum evolution induced by a time-periodic Lindbladian, $\mathcal{L}(t)$,~$\mathcal{L}(t+T)= \mathcal{L}(t)$~\cite{X3}.
The reason for that is that a generator -- if it is of the Lindblad form -- has to fulfill some very specific properties. Currently, the problem of the existence  of effective Floquet-Lindbladians is actively discussed in the literature~\cite{X4,X5,X6,X7,X8,X9}, and different expansion techniques are used to derive Floquet-Lindbladians. 
However, most of the  illustrative results are limited by $N=2$.

One of the current trends in the quantum community is to address many-body systems. In the context of the Floquet-Lindbladian problem that means that it is important to step beyond single spin/qubit models.  Here we approach the  problem from a new perspective, by considering it as a generic optimization problem and trying to analyze it by using the toolbox of Machine Learning methods. 
Our motivation is that some of these methods may give us a chance to get beyond the limit set by the reduced problem.
\end{quotation}


\section{Introduction\label{sec:1}}
We start right with the problem formulation and consider the time-dependent Markovian master equation \cite{Breuer2004,BreuerEtAl09}
\begin{align}
\label{eq:tdm}
\dot{\rho} = \mathcal{L}(t)\rho = -\frac{i}{\hbar}[H(t),\rho] + \mathcal{D}(t)\rho,
\end{align}
governing density operator $\rho$ of the model system. The evolution is set by a time-periodic
generator of the  Gorini-Kossakowski-Sudarshan-Lindblad form (henceforth "Lindbladian")~\cite{gorini,lindblad},
$\mathcal{L}(t)=\mathcal{L}(t+T)$. It is characterized by a  time-periodic
Hamiltonian $H(t)$ and the dissipative part 
\begin{align}
\mathcal{D}(t)\rho = 
\sum_i \gamma_i(t) \big[L_i(t)\rho L_i^\dag(t)-\frac{1}{2}\{L_i^\dag(t)L_i(t),\rho\}\big],
\end{align}
with jump operators $L_i(t)$ and non-negative rates $\gamma_i(t)$. In general, jump operators and rates are also time-periodic, with the same period $T$.
Under these conditions it is guaranteed that, for any time $t$, the corresponding  evolution can be reproduced  with a completely positive (CP) and trace preserving (TP) map~\cite{BreuerEtAl09},
\begin{align} 
\mathcal{P}(t)=\mathcal{T}\exp\left(\int_0^t\mathrm{d}t\mathcal{L}(t)\right).
\end{align}
where $\mathcal{T}$ is the standard time-ordering operator.

We will address the stroboscopic evolution set  by the map 
\begin{align}
\label{eq:oceo}
\mathcal{P}(T)=\mathcal{T}\exp\bigg[\int_0^T\!\rd t\, \mathcal{L}(t) \bigg].
\end{align}
The repeating action of this map induces the time-discrete evolution of the system,  $\rho(0)$ we have $\rho(nT) = \mathcal{P}(T)^n \rho(0)$. We now define a Floquet generator as a time-independent superoperator $\mathcal{K}$, such that
\begin{align}
\mathcal{P}(T) =  \exp\left({\mathcal{K} T}\right) \quad\text{  or   }\quad  \mathcal{K} =\frac{ \log(\mathcal{P})}{T} 
\label{eq:generator-cand}
\end{align}
for the open driven system described by  Eq.~\eqref{eq:tdm}.
If the Floquet generator can be recast into the Lindblad form, i.e., in the form given by  Eq.~(1) but with all operators and rates time-independent,  we  call the corresponding Floquet generator  \emph{Floquet-Lindbladian} and write
\begin{equation}
\mathcal{L}_F = \mathcal{K}.
\end{equation}
Now we pose the main question: Is there a Floquet-Lindbladian $\mathcal{L}_F$ for a given time-periodic Lindbladian $\mathcal{L}(t)$,
$\mathcal{L}(t+T) = \mathcal{L}(t)$?

It was demonstrated \cite{X1,X2} that the answer to another question "Is there a Lindbladian for the given completely positive trace-preserving (CPTP) map?" [which is the map $\mathcal{P}(T)$ in our case] can be obtained by reducing the original problem to an integer nonlinear programming problem. Therefore, formally, there is a way to get the answer to the question about the existence of Floquet-Lindbladian.

The number of variables $n$ in the reduced problem can be estimated straightforwardly. It is limited by the maximal possible number  of complex conjugated pairs in the spectrum of the map~\cite{X1,X2},  $n_{\mathrm{max}} = \lfloor \frac{N^2-1}{2}\rfloor$ (one is subtracted because at least one of the eigenvalues is equal to $1$). Here $N$ is the dimension of the Hilbert space $\mathcal{H}$ the map is acting in and $\lfloor...\rfloor$ is the floor operation, i.e., it gives the greatest integer less than or equal to the real input. For example, for $N=8$ we could have  up to $n=31$ integer variables  and the answer to the question is "yes" if there is  at least one  point of the lattice $\mathbb{Z}^{31}$ for which the set of the necessary (and altogether sufficient) conditions~\cite{X1} is fulfilled. 

It is indeed not possible to check all the  points -- even of $\mathbb{Z}^1$ -- if the lattice is not limited. Luckily, a convex feasible region in the real space $\mathbf{R}^n$ can always be out-shaped~\cite{X1}. To prove that the answer is "yes" or "no", we have to check all integer points inside the feasible region (or use the Khachiyan-Porkolab algorithm~\cite{Ellipsoid} instead). It was shown that this  problem can be reduced to a well-known NP-complete -- with respect to $n$ -- problem~\cite{X2}. That means that any practical algorithm based on this reduction will have an exponential complexity with respect to $n$.

The answer can be  easily obtained for a one-qubit ($N=2$) model because in this case $n=1$ and the feasible region is just a finite interval~\cite{X2,X3,X8}. However, already for two-qubit models, the complete check of integer points inside a $7$-dimensional volume can take a substantial time. We implemented an algorithm~\cite{unpub} that allowed us to out-shape the feasible volume and test it with several popular multi-spin models. It turned out that even in the case  of three spins  we have to deal with up to $10^9$ integer points for some, physically-relevant, values of model parameters. Even by taking into account the embarrassingly parallelizable character of the task (each point can be checked independently) and by running the algorithm on a medium-size cluster, one would need to wait for several hours before getting the answer~\cite{unpub}.

It is therefore doubtful that, by following this path, we would be able to get answers for four-qubit models in the general case, simply because it would not be possible to check all the integer points inside the feasible volume in the $127$-dimensional space. Yet here we need to recall that the algorithm we have used so far is based on a reduction of the original problem to a known NP-complete problem. Strictly speaking, the fact that one problem can be reduced to another, however, does not mean that both are equally complex. To speculate a bit further, many specific properties of the original matrices involved in the problem formulation were neglected in the course of the reduction so there is  still a chance that the original problem, even though still belonging formally to the same NP-complete class, can be solved faster and with less computation resources~\cite{parameter}.

After realizing the complexity of the problem, it is very natural to think about alternatives, which could allow us to solve problems for a reasonable large number of qubits -- at the expense of obtaining an answer which is not always correct but correct with some reasonably  high accuracy. Here Machine Learning (ML) methods look like immediate candidates~\cite{rmp2019}.  We made the first step in this direction in Ref.~\cite{X4}, where we tried to implement ML methods used for computer vision problems to reconstruct the boundary between 'yes/no' region on the parameter plane of  a two-qubit model. Unfortunately, the potential of this approach is rather limited as it requires extensive calculations on a coarse grid in the parameter plane for every new model and, therefore, is not capable of generalization. In addition, by training networks on images of 'yes/no'-boundaries we do not get closer to the understanding of mechanisms that determine the existence  (or non-existence) of Floquet-Lindbladians.

In this paper, we use several ML algorithms and implement a parameterization which is based on specific properties of the operators and matrices related to the original problem. By using a one-qubit model, for which the correct answer can easily be obtained~\cite{X1,X4}, we demonstrate that  ML methods can "learn" the 'yes/no'-partition of the parameter space. We also discuss what methods and parameterizations give the best accuracy. 

\section{Reduced problem \label{sec:2}}

In order to be a Lindbladian, operator $\mathcal{K}$, Eq.~(5), has to fulfilled two conditions. First, it has to  preserve Hermiticty. In case $\mathcal{K}$ is a logarithm of a Floquet map, this is guaranteed since the map itself is Hermiticity preserving. Next, $\mathcal{K}$ has to be \emph{conditionally} completely positive \cite{X1}. Formally, this means that 
\begin{align}
\Phi_{\perp} \mathcal{K}^\Gamma\Phi_{\perp} \geq 0,
\label{eq:test-cond-comp}
\end{align}
where $\Phi_{\perp} = \id - \ket{\Phi}\bra{\Phi}$ is the projector on the orthogonal complement of the maximally entangled state 
$\ket{\Phi}=\sum_{i=1}^N \left(\ket{i} \otimes \ket{i}\right)/\sqrt{N}$ with $\lbrace\ket{i}\rbrace$ denoting the canonical basis of ${\mathcal{H}}$. 

A new object, $\mathbf{C} = \mathcal{K}^\Gamma = N (\mathcal{K} \otimes \id)[\ket{\Phi}\bra{\Phi}]$, is the \emph{Choi operator}~\cite{choi} corresponding to $\mathcal{K}$ which acts in the product Hilbert space $\mathcal{H}^2$. If $\mathcal{K}$ is given in the matrix form (by using some basis)
\begin{equation}\label{super2}
   \widehat{\mathcal{K}}_{ij,kl} := \langle i \otimes j|  \mathcal{K} | k \otimes l\rangle .
\end{equation}
the matrix form of the corresponding Choi operator is related to it by the reshuffling operation~\cite{duala}
\begin{equation}\label{resh}
\mathbf{C} = \widehat{\mathcal{K}}^\mathcal{R},~~~ \mathbf{C}_{ij,kl}  = 
   \widehat{\mathcal{K}}_{ik,jl}.
\end{equation}
The reshuffling operation $\mathcal{R}$ is an involution so that being repeated twice it results in the identity transform. In the case when $\mathcal{K}$
is a CPTP map, the corresponding Choi operator is a state~\cite{choi,duala}, i.e., a density operator (though not normalized), in $\mathcal{H}^2$. For example, for any Floquet map $\mathcal{P}(T)$ we could get the corresponding state.

Condition (7) can be recast in some matrix inequality~\cite{X1}, if we use spectral decomposition of $\mathcal{K}$. Since the Floquet map is Hermiticty preserving, 
its spectrum is invariant under the complex  
conjugation. Therefore, the corresponding  $N^2$ eigenvalues are either real or appear as complex conjugated pairs (strictly speaking, the spectrum could also include eigenvalues of odd degeneracy). We denote 
the numbers of real eigenvalues as $m$ and complex pairs as $n$. 
$\mathcal{P}(T)$ can be represented as
\begin{align}
\mathcal{P}(T) =  \sum_{r=1}^{m} \lambda_r P_r +  \sum_{c=1}^{n}  \left(\lambda_c P_c +  \lambda_c^* P_{c*}\right),
\label{eq:map-jordan-form}
\end{align}
where $\lambda_r$ are the real eigenvalues, $\lbrace \lambda_c, \lambda_c^*\rbrace$ are the pairs of complex eigenvalues, and $P_x$ the corresponding  projectors~\cite{X1}.

Any logarithmic branch, Eq.~(5), of $\mathcal{P}(T)$ can be represented as
\begin{align}
\mathcal{K}_{ \{x_1,...,x_{n}\}} = \mathcal{K}_0+ i \omega \sum_{c=1}^{n}  x_c \left( P_c -   P_{c*}\right),
\label{eq:candidates-LF}
\end{align}
where $\mathcal{K}_0$ is the  principle branch and $\omega = \frac{2\pi}{T}$. 
Therefore, every branch is parametrized with $n$ integers, $\{x_1,...,x_{n}\}$, i.e., it corresponds to a vertex $\mathbf{x}=\{x_1,...,x_{n}\}$ of the $\mathbb{Z}^{n}$ lattice. Now we introduce a set of operators (that are, in fact, Hermitian)
\begin{equation}
	V_0 = \Phi_{\perp} \mathcal{K}_0^\Gamma\Phi_{\perp}, ~~ V_c = i \omega \Phi_{\perp} (P_c - P_c*)^\Gamma\Phi_{\perp}, ~~ c = 1, \dots, n.
\end{equation}
and arrive at the following test~\cite{X1}:

\noindent\fbox{%
    \parbox{0.47\textwidth}{%
Generator $\mathcal{K}_{\mathbf{x}}$ is Lindbladian iff there is  a set of $n$ integers, $\mathbf{x} \in \mathbb{Z}^{n}$, such that
\begin{equation}
V_{\mathbf{x}}  =  V_0 + \sum_{c=1}^{n} x_c V_c \geq 0.
\label{eq:Fl-matrix-cond}
\end{equation} 
    }%
}

Such type of matrix inequalities (and related programming problems) is  well known in the control theory; see. e.g., Ref.~\cite{control}. The crucial difference is that in there  vector $\mathbf{x}$ is considered to be  real. If matrices $V_c$ are real and symmetric, the 
inequality out-shapes a convex feasible region known as spectrahedron~\cite{spectra}. We deal with integers and this mere fact makes the problem NP-complete (with respect to the number of the matrices in the sum). 
However, we can benefit from 
the fact the feasible region  (13) is convex if $\mathbf{x} \in \mathbb{R}^{n} $. That is, we can limit the number of integer points needed to be checked. 
This was implemented in the algorithm we discussed in the introduction.

Finally, imagine that we can quantify distance from 'Lindbladianity' for any  generator $\mathcal{K}_{\mathbf{x}}$. Then, if the answer is "no" and  there is no such integer vector ${\mathbf{x}}$ that condition (13) holds (and, correspondingly, no Floquet-Lindbladian exists), we can define the distance from Markovianity for $\mathcal{P}(T)$  by picking the branch which gives the minimal distance. Such measure was proposed by Wolf \emph{et al.}~\cite{X1}. It is based on adding a noise term $\mathcal{N}$, that is the generator of the depolarizing channel, 
$\exp(T\mu \mathcal{N})\rho = e^{-\mu T} \rho + [1 - e^{-\mu T}] \frac{\mathbf{1}}{N}$, weighted with the strength $\mu$, to 
the generator and determining the minimal strength required to make at least one of the logarithmic branches Lindbladian , i.e.
\begin{align}
\mu_{\mathrm{min}}=\inf \left\lbrace \mu \geq 0 {\Big\vert} \begin{array}{l}   \mathcal{K}_{\mathbf{x}} + \mu \mathcal{N}  \text{ is a valid} \\
\text{Lindblad generator}\end{array} \right\rbrace.
\end{align}

\section{Methods\label{sec:3}}

We select eleven popular ML methods implemented in the scikit-learn library \cite{scikit-learn}. Namely, we use the k-nearest neighbors method (kNN) \cite{X5,X6}, support vector machine (SVM) with linear, polynomial and RBF kernels \cite{X7,X9}, decision trees \cite{X10,X11}, random forest (RF) \cite{X12,X13}, fully connected neural networks (FNN) \cite{X14,X15}, AdaBoost \cite{X17}, linear (LDA) and quadratic (QDA) discriminant analysis \cite{X18}, and naive Bayes classifiers \cite{X19}. All these methods are well described in the literature, so in this section we only explain the reasoning behind our choice.

The performance of   kNN , decision trees, and random forest methods
depends on how the elements of the training and test samples are distributed in the space. In all these methods, a direct comparison of an element from a test sample with a training sample and search for a similar element, are used. We could expect  good performance of the methods in data interpolation if the training sample is representative, but the possibility of  equally good extrapolation is questionable. Various implementations of SVM, discriminant analysis, and naive Bayes classifier have different features. Thus, we may expect to get reasonably accurate functional models of the boundaries between the 'yes' and 'no' regions. 

The random forest, fully connected neural networks, and AdaBoost methods  are able to build very complex models and highlight non-obvious features and can solve a complex problem in an accurate way. We are not only motivated by the idea to demonstrate the potential of ML methods in getting the answer for single-qubit models. In fact, we hope to get an intuitive understanding of how the specific mathematical nature of the problem influences the results of the classification. We expect that by analyzing the performance of these methods we can get an additional insight.

All the selected methods have a number of meta-parameters, some of which are set by default, while others are selected and optimized  to get the best results. In particular, we vary the following parameters: the number of neighbors in kNN (typically 3-7), the degree of the kernel polynomial in the polynomial SVM (typically 2-11), the height of the decision tree (typically 3-15), the height (2-9) and the number (typically 50-300) of trees in random forest, the number of layers (typically 3-4) and their size (from 8 to 512) in fully connected neural networks, as well as the type of the AdaBoost classifiers (short decision trees) and the number of these classifiers (typically 50-300). In the following sections, we present only the best results obtained for each of the methods.

\section{Model\label{sec:4}}

As a test bed we  used a single spin(qubit) model described with the following Hamiltonian~\cite{X3} (Problem~I):
\begin{equation}
H(t)= \frac{\Delta}{2} \sigma_z + E cos(\omega t+\varphi) \sigma_x,~~ L = 
\sigma\_
\end{equation}

We also modified this model by adding one extra operator term to the Hamiltonian (Problem~II):
\begin{equation}
H(t)=  \frac{\Delta}{2}(\sigma_z+\sigma_y)+E cos(\omega t+\varphi) \sigma_x, ~~ L = 
\sigma\_
\end{equation}

For both models, we use the following parameter values: $\Delta = 1, \gamma = 0.01$. Such parameters as amplitude $E$, frequency $\omega$, and phase shift $\varphi$, are varied to generate data sets for the ML algorithms.

As it is shown in the previous section, solution  of the Floquet-Lindbladian existence problem reduces to the analysis of the properties of the logarithm of Floquet map, $\mathcal{P}(T)=\mathcal{T}\exp\bigg[\int_0^T\!\rd t\, \mathcal{L}(t) \bigg]$. This Floquet map, in turn, can be transformed  into a state by using the Choi–Jamio\l{}kowski isomorphism~\cite{choi,duala}. Represented as a matrix, it is positive semidefinite and has trace $N$. 
The isomorphism provides a natural way to parametrize Floquet maps in order to produce inputs for the ML algorithms. We will consider three possible encoding schemes that are to use as inputs: (i) the eigenvalues of Choi matrices, (ii) whole Choi matrices, and (iii) eigenvalues and eigenvectors of Choi matrices.  


\section{Results\label{sec:5}}
\subsection{Methodology}

We start with a brief summary of the results presented in this section.

We consider two one-qubit models, Problem~I and Problem~II, defined in the previous section).
Both are parametrized with two parameters, amplitude $E$ and  frequency $\omega$ of the modulations.
We start with Problem~I and analyze the performance of the selected  ML methods (Section III) in providing the answer to the question "Is there a Floquet-Lindbladian?", for different regions on the $\lbrace E,\omega \rbrace$ plane. Instead of the original Floquet maps, we use the corresponding Choi matrices as the main source of information for classification. After obtaining  encouraging results for Problem~I, that is separability in the space of the eigenvalues of the Choi matrices, we check  whether the obtained classifiers are capable of coping with Problem~II.
After realizing that the quality of the results became significantly worse, we consider alternative  methods of parametrization, still based on Choi matrices. We find that one of these parametrizations turns to be the best.

For Problem~I, we generate samples by varying phase shifts $\varphi$. For each value of $\varphi$, we go through the values of the amplitude $E$, from 0 to $\pi$, and the frequency $\omega$, from 0 to $2\pi$ with a step of $\pi/25$. Each dataset corresponding to phase shifts $\varphi = \lbrace 0, \pi / 4, \pi / 3, \pi / 2, 2\pi / 3, 3\pi / 4 \rbrace$ is randomly divided into training and validation samples in the proportion of $90\%$ and $10\%$, respectively. A test sample is organized in a similar way, but for shifts $\varphi = \pi/8$ and $\varphi = 5\pi/8$. Further, similar samples are generated for Problem~II.

We use three conventional metrics: accuracy, f1-score, and Area Under the Curve (AUC). The first metric, accuracy, is the percentage of correct answers for the presented data. This metric is not always fair, since in many problems it is not easy to balance the number of objects of different classes in the samples. In our case, f.e., about $70\%$ of the data belongs to the class of problems with the answer 'yes', so even an elementary classifier that always answers 'yes' will receive an acceptable accuracy of $\sim 0.7$. Therefore, we also use two other metrics to take care of the imbalance in the datasets.

The f1-score \cite{X21}  is the harmonic average between precision and recall. This metric is a compromise between the two, where precision is the ratio of true-positive results to all positive results, and recall is the ratio of true-positive elements to all true-positive and false-negative objects in a sample. The need for a compromise is due to the fact that often we can increase precision, decreasing recall, or increase recall, but decrease precision. The third metric, AUC~\cite{X22}, is the integral indicator which estimates the quality of a classifier and its errors. It is not always possible to use it independent of the context, since this indicator is very sensitive to noise, but could be informative as a complement to other metrics.

\subsection{Results}
\subsubsection{Classification based on eigenvalues of Choi matrices}
Feature extraction plays one of the key roles in solving classification problems using ML methods. Relevant feature extraction affects the accuracy of classifiers, allowing ML methods to discard insignificant information and focus on the main features. During the last decade, substantial progress in this direction has been made when solving a wide range of problems from various fields of science and technology. These fields include  computer vision, image processing, computational biomedicine, natural language processing, and other applications. However, for the problems from such field as computational physics, there are no well-defined guidelines, and intuition still plays an essential role. Intuitive guesses are often supplemented by several trial-and-error rounds which result in finding the feature space in which \textit{separability} is sharp enough.

In the case of the Floquet-Lindbladian problem, our intuition tells that the properties of the Choi matrices of the Floquet maps can open a way to the desired separability. It is known that the Choi matrix of the quantum map reflects the complete positivity (or its absence) of the latter in a very transparent way: The map is completely positive if its Choi matrix is positively semidefinite, i.e. all the Choi eigenvalues are non-negative~\cite{duala}. Therefore, it is natural to assume that the property of being Markovian (or the absence of such property) of the given CPTP map can be encoded in the corresponding Choi matrix. However, what properties of the Choi matrix to use and what parametrization is optimal are open questions. We start with the eigenvalues of the Choi matrices as features to classify the 'yes' and 'no' answers. 


\begin{figure*}[h]
\includegraphics[width=1\textwidth]{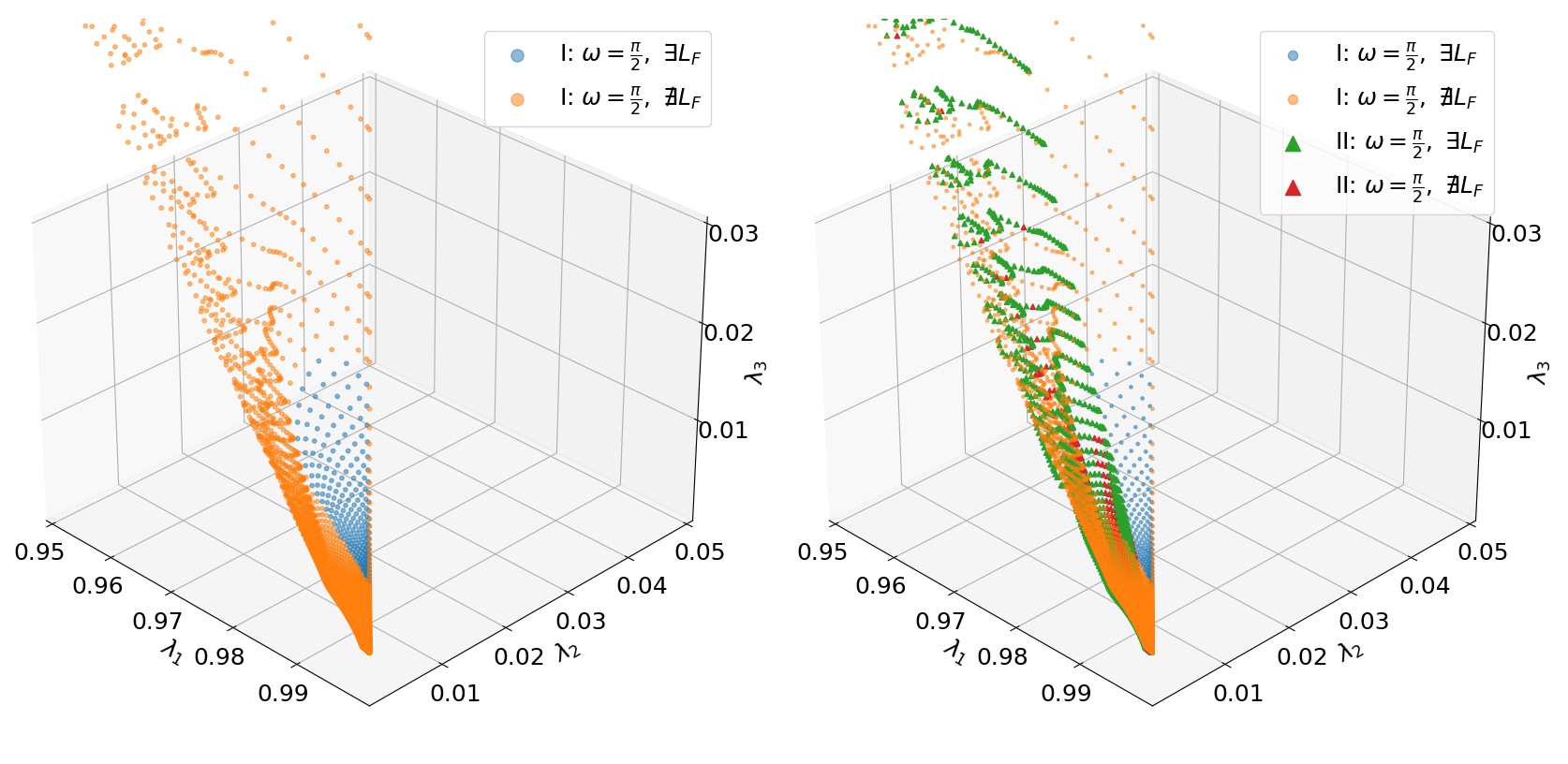}
\caption{Distribution of 'yes' (orange and red dots) and 'no' (blue and green dots) answers in the space of the three largest eigenvalues of the Choi matrices of the Floquet maps. Left: distribution for Problem~I, Eq.~(15), for phase shift of $\pi/2$. Right: Distributions for Problem~I, Eq.~(15), and Problem~II, Eq.~(16), plotted together. Classification is expected to be more complicated in this case because the separability cannot be  detectable visually, in  contrast to the distribution for Problem I alone (left panel).} \label{fig:1}
\end{figure*}

The eigenvalues of the Choi matrices are real positive numbers which sum up to $N$. The space of eigenvalues in the one-qubit problem is a three-dimensional positive manifold in the four-dimensional real space. The possibility to visualize the eigenvalues in 2D space and thus to examine the position of the points corresponding to the 'yes' and 'no' answers is an important advantage of this parametrization. To test  this idea, we consider Problems I and II and vary values of the amplitude and frequency for a specific fixed value of the phase shift. 

For every parameter set, we construct the Choi matrix and calculate its eigenvalues numerically. Further, by using the test (13), the answer to the question of the existence of Floquet-Lindbladian is obtained. The generated dataset consists of lists of eigenvalues, labeled with "0" if the answer is 'no' and "1" if the answer is 'yes'. Fig.~\ref{fig:1} (left) shows that the distribution of the answers is not chaotic but exhibits some structure so potentially it can be partitioned into 'yes/no' sub-manifolds. We find this behavior typical, and the partition is present for both models, I and II, when they are considered separately. 
However, as it is discussed in the next section, if we plot the 'yes/no' points for both models, without differentiating between them, we get a cloud of points which no longer exhibits signatures of partition. 

\begin{figure*}[h]
\includegraphics[width=0.8\textwidth]{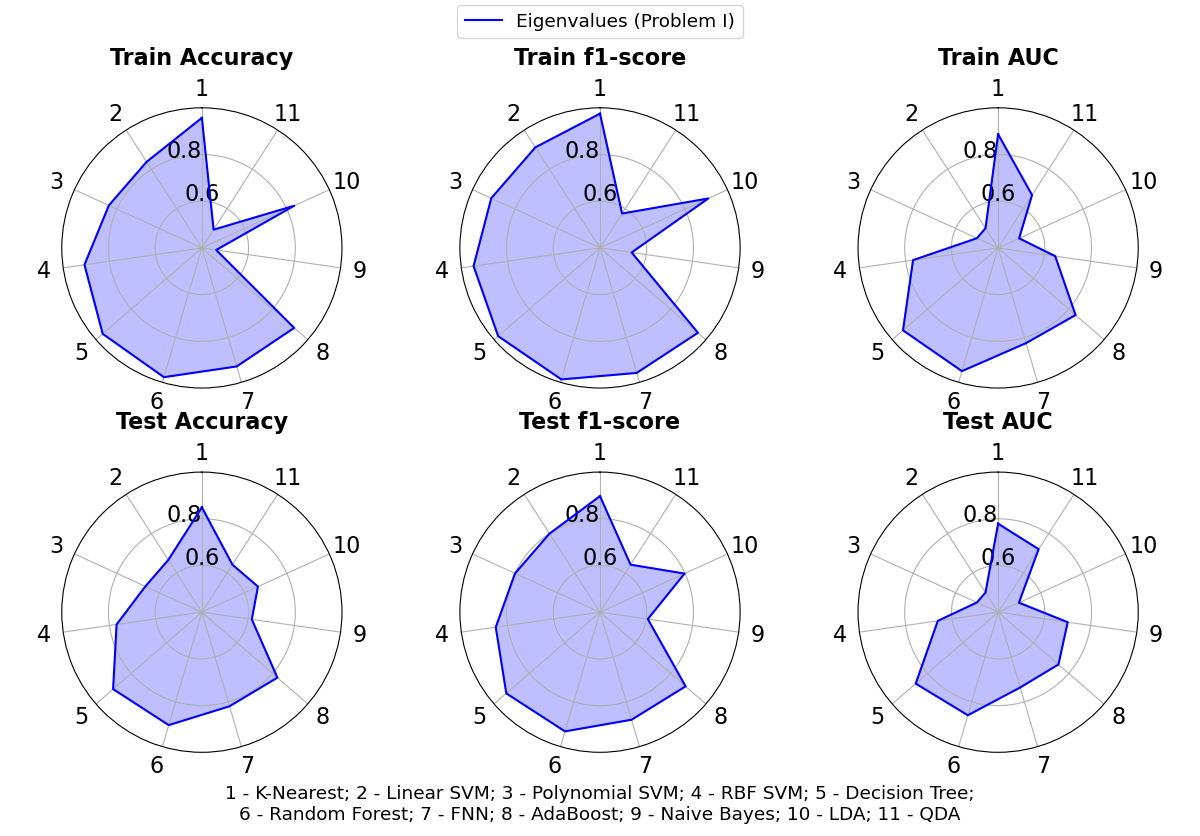}
\caption{Classification accuracy for datasets of Choi eigenvalues of Problem I, for training (top row) and test (bottom row) samples. The accuracy is quantified by using three different measures (see Section VA). The center of a diagram corresponds to accuracy $0.4$.} \label{fig:2}
\end{figure*}

Next, we built classifiers using eleven selected ML methods. The results for Problem~I are presented with Fig.~\ref{fig:2}. The main observation is that the Nearest Neighbors, Decision Tree, and Random Forest methods yield the best results (in terms of the metrics used) and significantly outperform other methods. The reason for this is that  these methods  compare the classified point of the parameter space with the original training sample, thereby taking into account the fact that the points are close in the sense of a particular metric. Considering by tuning the  amplitude, frequency, or phase shift we do not detect substantial changes (for both problems, I and II), we conclude that the classification scheme based on the Choi eigenvalues works reasonably well.

\begin{figure*}[h]
\includegraphics[width=0.8\textwidth]{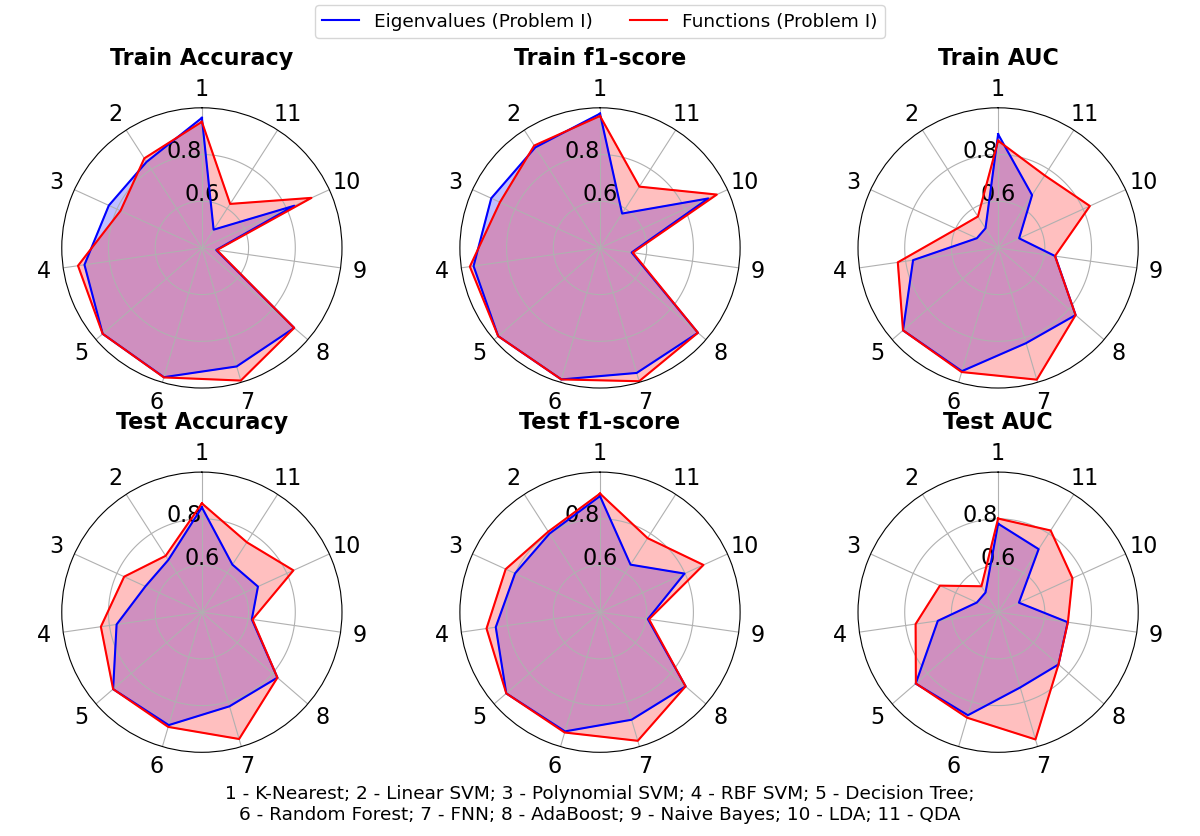}
\caption{Classification accuracy for datasets of Choi eigenvalues (blue) and of their functions, square roots and and fourth-order roots (red). Datasets were generated for Problem I. The accuracy is quantified by using three different measures (see Section VA). The center of a diagram corresponds to accuracy $0.4$.} \label{fig:3}
\end{figure*}

Is it possible to improve the classification? To get an insight, we expand the feature space by adding various functions of the Choi eigenvalues such as logarithms, exponents, degrees, roots, trigonometric, hyperbolic functions, and their combinations. We assume that in the extended space, methods would be able to detect linear or other, relatively simple, separability. We obtain the best separability in the  space consisting of eigenvalues in powers of 1, $1/2$, and $1/4$; see Fig.~\ref{fig:3}. 
It is noteworthy that those methods that previously worked poorly have significantly improved their results. The fully connected neural network outperforms  other methods in the sense  of the classification accuracy, being able to accurately approximate the separating surface in the feature space.

The next important question  is to what extent the developed classifiers are capable of generalization. Note that so far they were trained to solve one specific problem (Problem~I or Problem~II), albeit with different  parameters. Now we want to check how the classifiers developed for Problem~I can cope with Problem~II.

\begin{figure*}[h]
\includegraphics[width=0.8\textwidth]{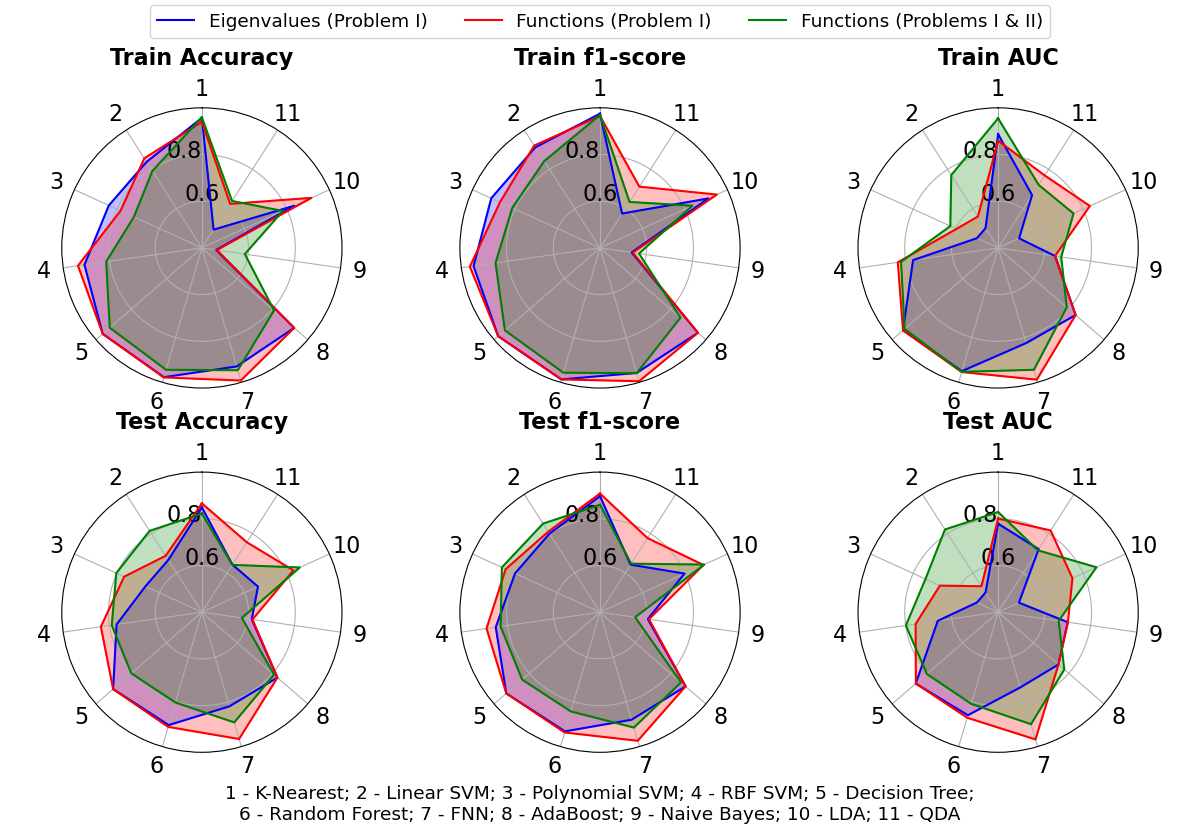}
\caption{
Classification accuracy for datasets of Choi eigenvalues (blue) and of their functions,  square roots and fourth-order roots (red). Datasets were generated for Problem I. Additionally, we train and test models with mixed datasets consisting of the data for both problems, I and II (green). The accuracy is quantified by using three different measures (see Section VA). The center of a diagram corresponds to accuracy $0.4$.} \label{fig:4}
\end{figure*}

Unfortunately, the outcome is disappointing. It turns out that the models trained and validated for Problem~I is not suitable for Problem~II. One could say that this is expected since all methods have not seen the data obtained for Problem~II during their training. However, the problems are  similar and, therefore, we naturally expect the models should be able to generalize. To improve the performance, we add data obtained for Problem~II to the training set. This dramatically improved the results; see Fig.~\ref{fig:4}.

However, it is still not possible to achieve results that compete with the metrics obtained during training and validating of classifiers on a single problem. The reason for this  can be understood by looking at Fig.~\ref{fig:1} (right). The presented plot shows that it is difficult to reach separability because when the samples from two problems are combined, some points appear surrounded by a cloud of points with the opposite answer. We do not present the feature space extended by various functions of the eigenvalues of the Choi matrices for this case, but obtained results make it clear that the corresponding parameterization does not lead to a better accuracy. Therefore, we should look for alternative parameterizations.

\subsubsection{Classification based on the elements of Choi matrices}

\begin{figure*}[t]
  \includegraphics[width=0.7\textwidth]{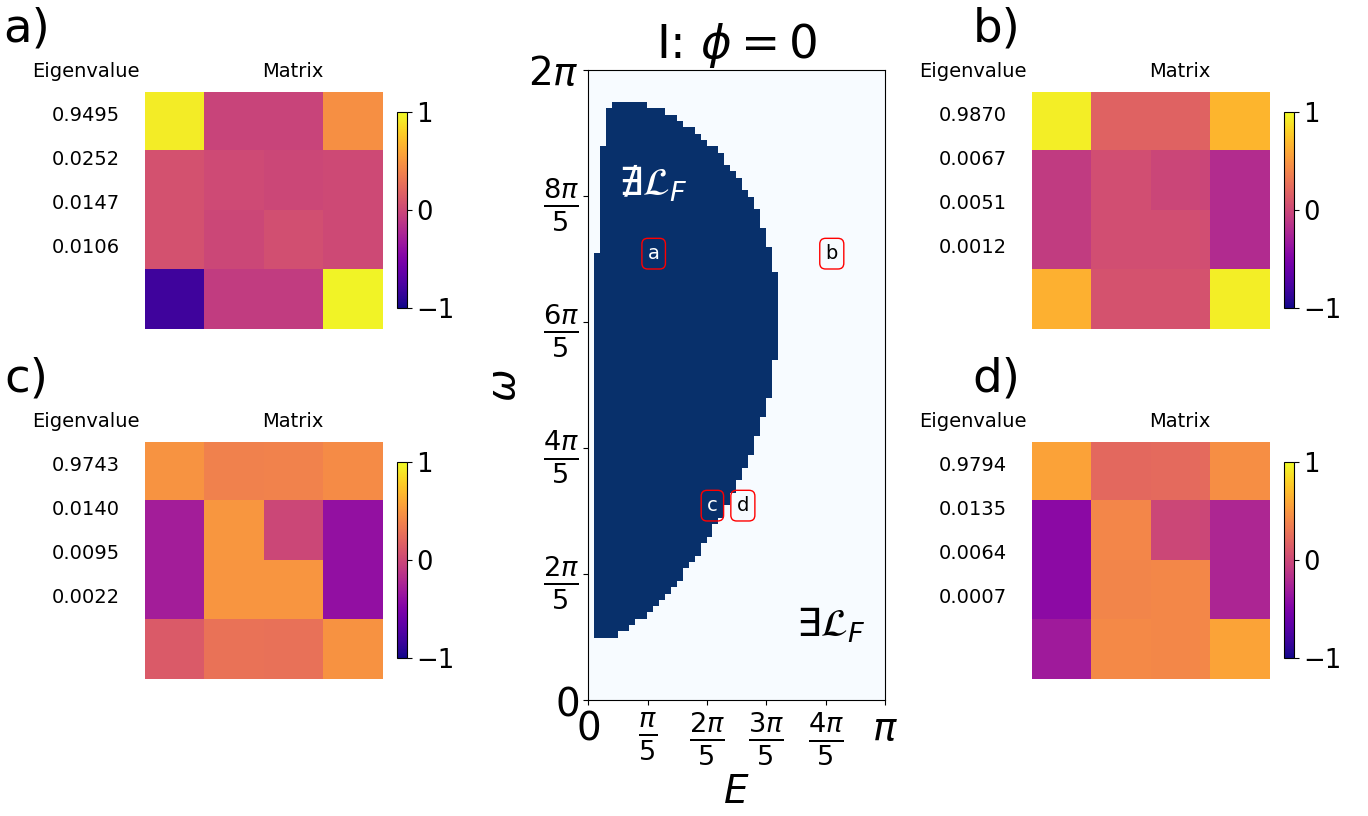}
  \caption{Parametrization of the Choi matrix.
   The upper half of the matrix corresponds to the real part, the lower one -- to the imaginary part of the matrix. Center: Diagram of 'yes' (light blue) and 'no' (dark blue) answers to the question of the existence of Floquet-Lindbladian for the Problem~I. Four combinations of parameters are used (a, b, c, d). Their positions on the $(E,\omega)$ plane are indicated on the central diagram (red squares). } \label{fig:5}
\end{figure*}

As we figured out in the previous section, the parametrization based on the eigenvalues of Choi matrices does not provide the possibility to generalize from one physical model to another model even if we add data from both models in the training dataset. Apparently, this is because such parameterization is rather reduced and we are losing too much information when condensing  properties of the original Floquet map into the corresponding Choi eigenvalues. Here we try an alternative parametrization based on the elements of the Choi matrix. Namely, we use the following parameterization: the upper triangle of the matrix, including the diagonal, containing the real parts of the elements, and the lower triangle containing the imaginary parts of its elements. Such a representation is complete, since we can unambiguously restore the original data, and all the elements of the constructed matrix are real numbers. Next, we vectorize the real matrix, by writing its elements one after another column-wise. Fig.~\ref{fig:5} presents examples of such parameterizations for four different parameters of Problem~I (Figs.~5a-d). 

\begin{figure*}[t]
\includegraphics[width=0.8\textwidth]{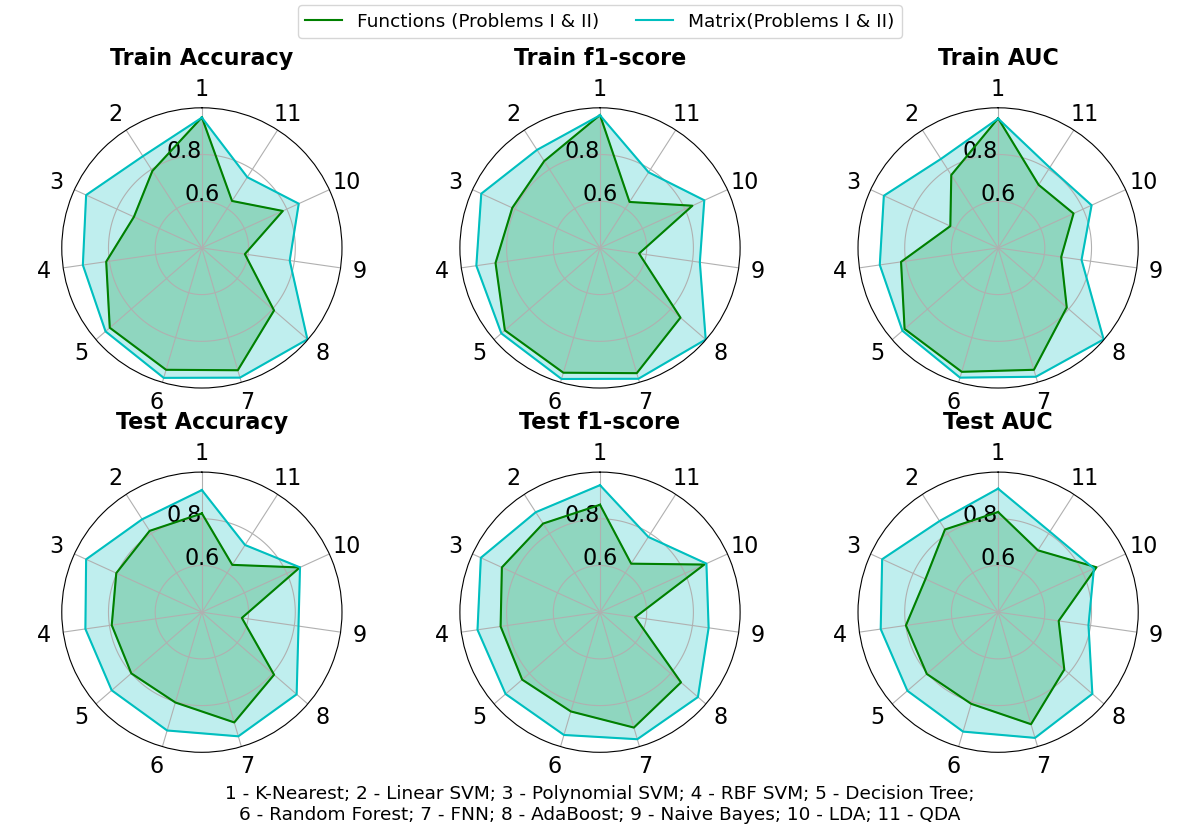}
\caption{Classification accuracy for datasets of the Choi eigenvalues, and their functions, square roots and fourth-order roots (green) compared to the accuracy obtained by employing datasets generated by using elements of the Choi matrices (light blue). Samples from Problem~I and Problem~II were used for training and testing. The accuracy is quantified by using three different measures (see Section VA). The center of a diagram corresponds to accuracy $0.4$.} \label{fig:6}
\end{figure*}

Using this parametrization, it is also possible to solve the classification problem by combining data from problems I and II both in the training and in the test samples. The best achieved results are summarized in Fig.~\ref{fig:6}. It turns out that almost all methods substantially improved their performance, both on the training sample and the test sample. AdaBoost, Neural Network, and Random Forest methods show very good results on the training set (errors less than $2.5\%$), while on the test set, the results are expected to be slightly worse in terms of accuracy.

\begin{figure*}[h]
  \includegraphics[width=0.8\textwidth]{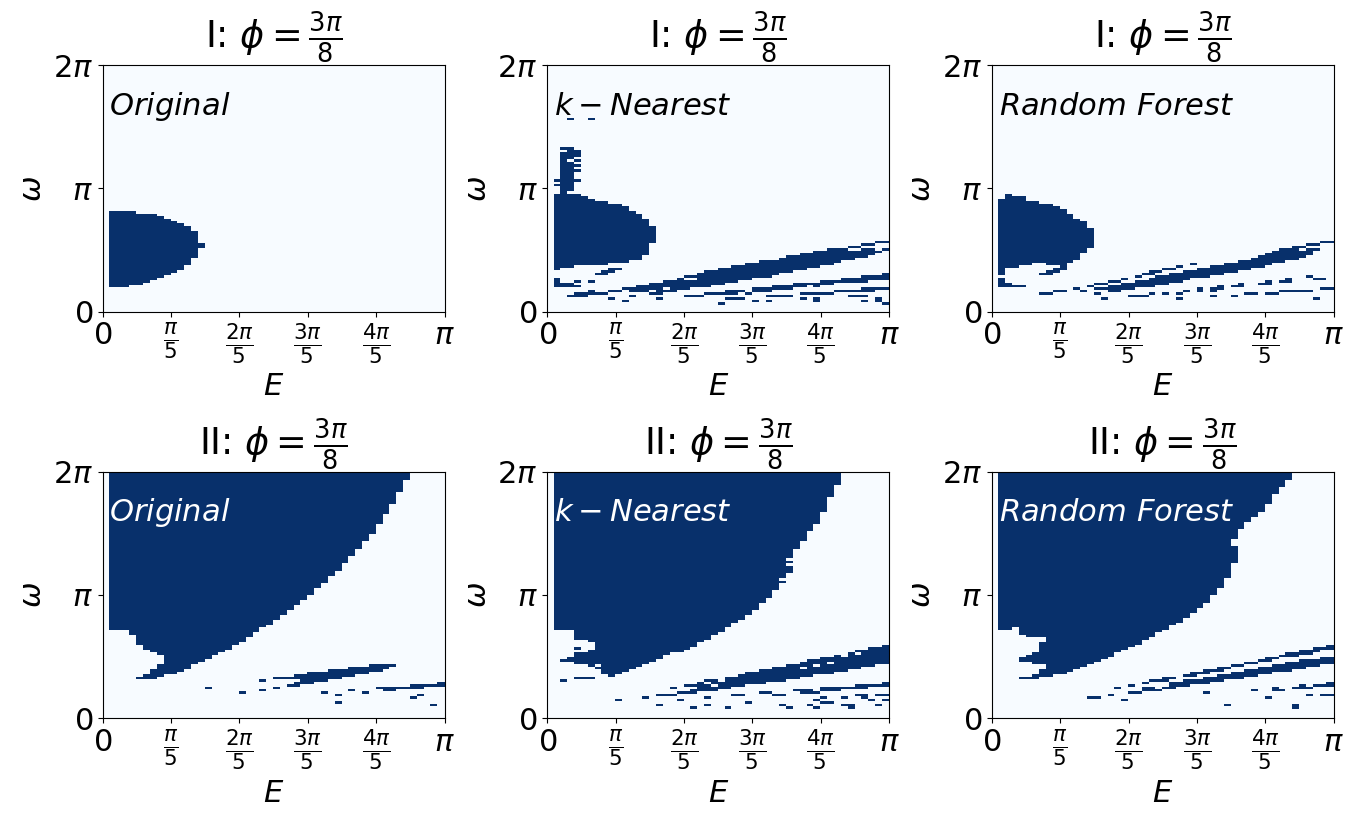}
  \caption{'Yes/no' partition of 
  the parameter space for Problems I (top row) and II (bottom row) for different values of phase shift. Partitions are obtained by using the test, Eq.~(13)[left] and ML methods,  the Nearest neighbors method [center] and the Random forest method [right]. For the ML methods the parametrization based on the elements of Choi matrices is used.} \label{fig:7}
\end{figure*}

To get more insight, we inspect the two-dimensional diagram on Fig.~\ref{fig:7} in order to understand when the classifiers are wrong. It is not correct to think that errors are localized at the vicinity of the border between the 'yes/no' areas. The errors of this sort are expected; however, we also find  error zones that are far from the border. For example, ML methods mistakenly detect extensive dark blue ‘petals’ corresponding to the answers 'no' when solving Problem~I. It is noteworthy that such ‘petals’ exist in Problem~II. Obviously, having learned such ‘petals’ in one of the problems, ML methods fail to learn to recognize whether these petals are in other problems or not. We tried to overcome this problem by balancing the contribution of data from several problems in the training sample, but this did not help. Moreover, we found that ‘petals’ appear erroneously even if we only use Problem~I for training and evaluation. Our conclusion is as follows: the considered parametrization is not suitable for an accurate feature selection. 

It seems that even though we are using all the information encoded  in the Choi matrix, the resulting approach does not lead to a better performance. The tendency is that simple models cannot generalize data from different problems and complex models are often overfitted. We conclude that we should try to identify features that are important for classification and train models on these features.

\subsubsection{Classification based on eigenvalues and eigenvectors of Choi matrices}
Finally, We consider parameterization of a Choi matrix based on the full eigenset, which includes  eigenvalues and eigenvectors. The matrix of eigenvectors is an $N^2 \times N^2$ orthonormal complex matrix. It is not very convenient to use it in this form since the values of its elements vary over a wide range. Given that angles are a key feature of eigenvectors, we decided to use spherical coordinates. To use this coordinate system, we have to unfold a complex vector of size $N^2$ into real vectors of the size of $2N^2$. In this case, it is possible to make the last imaginary coordinate (the last coordinate of the real vector) equal to 0. We converted such vectors into spherical coordinates as follows: 
\begin{equation*}
    \begin{gathered} 
r = \sqrt{x^2_1 + x^2_2 \dots x^2_{N^2-1} + x^2_{N^2}} \\ 
\varphi_1 =  \arccot  \frac{x_1}{\sqrt{x^2_{N^2} + x^2_{{N^2}-1} \dots x^2_3 + x^2_2}}\\ 
\varphi_2 =  \arccot  \ \frac{x_2}{\sqrt{x^2_{N^2} + x^2_{{N^2}-1} \dots x^2_4 + x^2_3}}\\
\dots \\ 
\varphi_{N^2-2} =  \arccot  \ \frac{x_{N^2-2}}{\sqrt{x^2_{N^2} + x^2_{N^2-1}}}\\ \varphi_{N^2-1} =  \arccot  \ \frac{x_{N^2-1} + \sqrt{x^2_{N^2} + x^2_{N^2-1}}}{x^2_{N^2}}
\end{gathered}
\end{equation*}

This approach has several advantages: 
\begin{enumerate}
  \item  There is no loss of information, that is, the original data (Floquet map) can be reconstructed.
  \item  The length of the vector is equal to 1 (and therefore is irrelevant). 
  \item  We can exclude  the coordinate of the angle $\varphi_{N^2-1} \in [0,2\pi]$ from the data, since it is equal to zero during computation for the zero component. Thus, the angles close to 0 and $2\pi$ are not close to each other in the generated data. 
  \item  All other angles $\varphi_i \in [0,\pi]$, and represent an ordinary hypercube.
\end{enumerate}

All angles were also normalized to the range $[0, 1]$ to get rid of unnecessary dependence on $\pi$.  

\begin{figure*}[h]
  \includegraphics[width=0.8\textwidth]{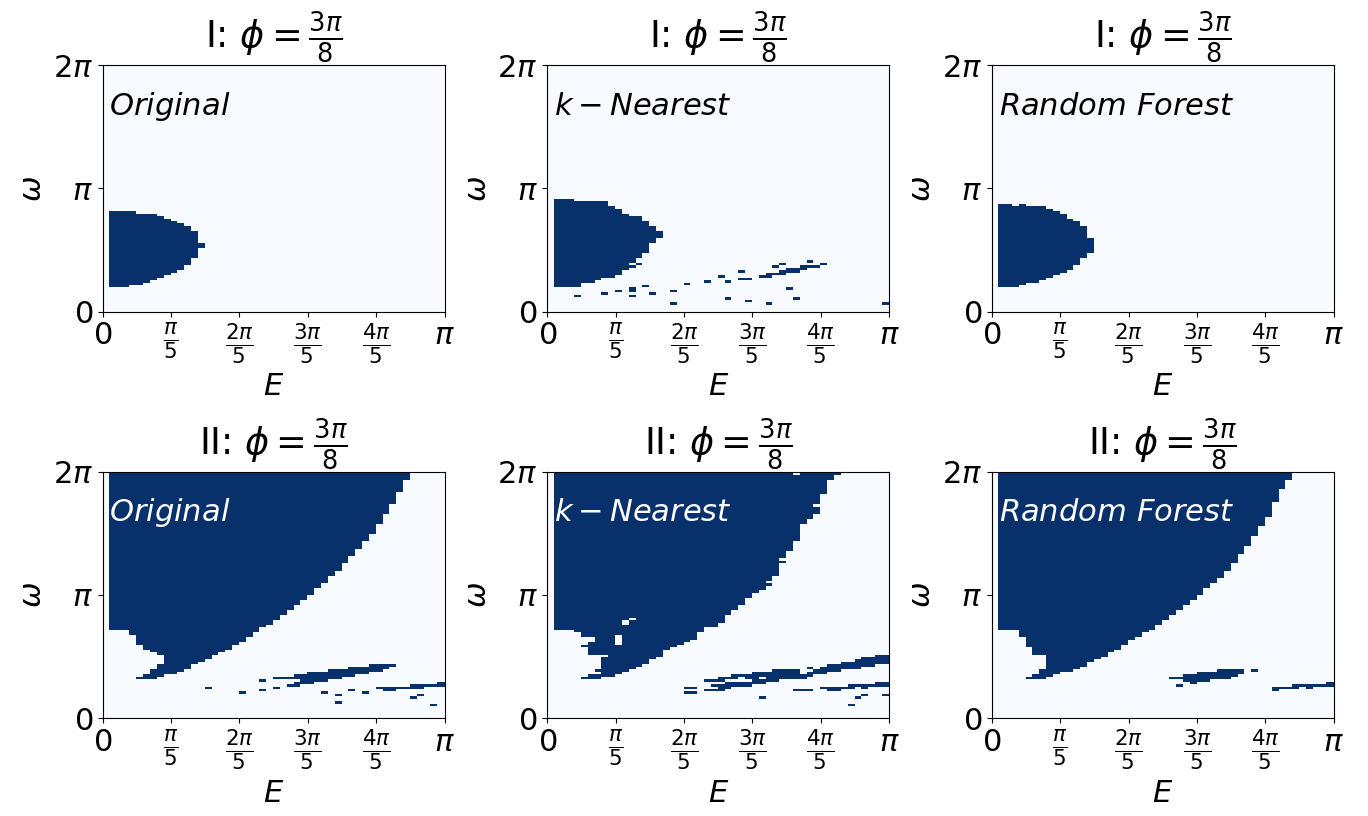}
  \caption{'Yes/no' partition of 
  the parameter space for Problems I (top row) and II (bottom row) for different values of phase shift. Partitions are obtained by using the test, Eq.~(13) [left] and ML methods, the Nearest neighbors method [center] and the Random forest method [right]. For the ML methods the parametrization based on the eigenvalues and eigenvectors of Choi matrices is used.} \label{fig:8}
\end{figure*}

\begin{figure*}[t]
  \includegraphics[width=0.8\textwidth]{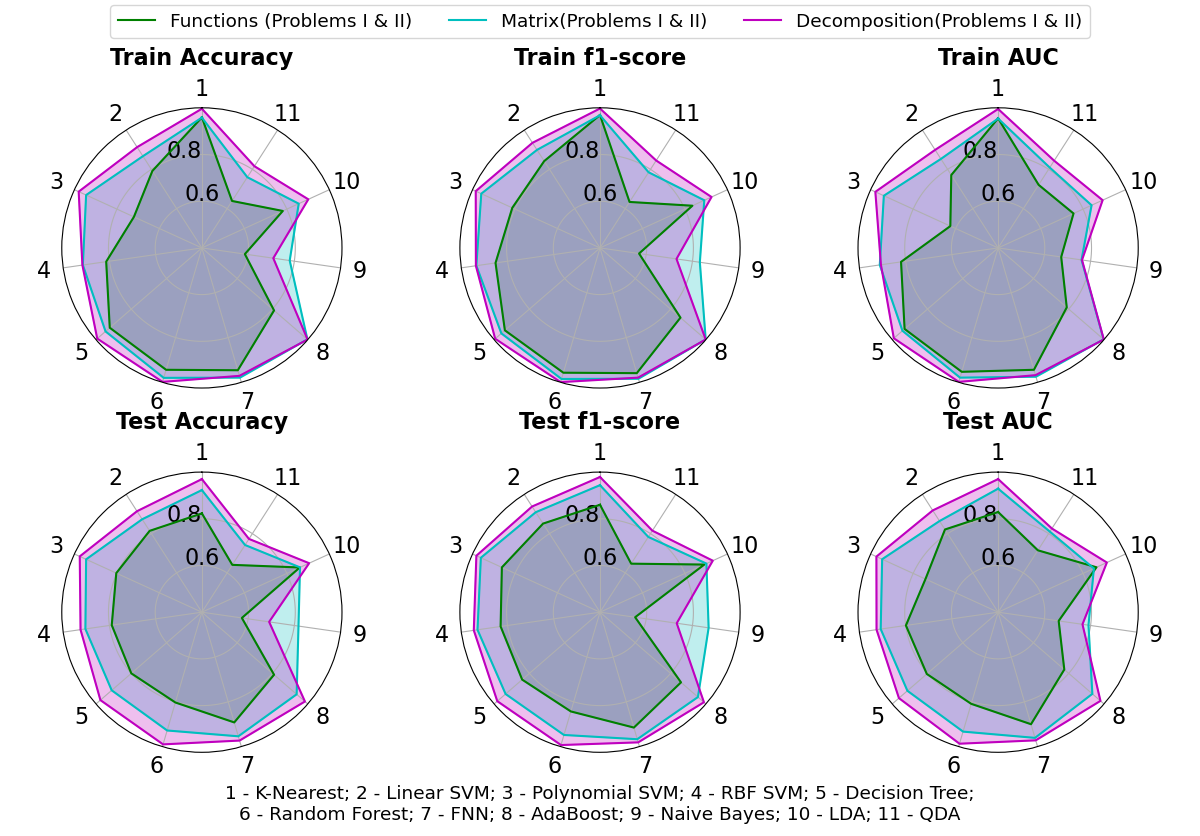}
  \caption{
  Classification accuracy for three datasets: a) generated by using the Choi eigenvalues and their functions, square roots and fourth-order roots (green); b) generated by using elements of the Choi matrices (light blue); c) generated by using the Choi eigenvalues and eigenvectors after appropriate normalization (magenta). Samples from Problem~I and Problem~II were used for training and testing. The accuracy is quantified by using three different measures (see Section VA). The center of a diagram corresponds to accuracy $0.4$.} \label{fig:9}
\end{figure*}

After constructing the classifiers, we implement the same training procedure as in the previous section, by using datasets obtained with both models, I and II. We find that the results are substantially better; see Fig.~\ref{fig:8}. However, there are still regions of wrong answers far from the 'yes/nor' border. As before, in some cases in Problem~I, the petals of answers 'no' were mistakenly found in those areas of parameters in which these petal-like regions are present in Problem~II. We analyze the datasets to  understand the origin of this behavior. We find that the adjacent points on the diagram are distinguished by a smooth change in the angles corresponding to real-valued coordinates, and by sharp changes in the angles corresponding to complex-valued coordinates. As a result, ML methods at some point begin to 'react' only to dramatic changes of the values, which leads to the decay of the accuracy of classification. To overcome this effect, we remove angles corresponding to the complex-valued coordinates from the datasets. After purging the data, we evidently are no longer able to reconstruct the original matrix, however we 
expect that this purge could improve the accuracy of classification. Fig.~\ref{fig:9} reports  the obtained results. 

First of all, the accuracy of every classifier has improved. Most of the methods now give correct answers, both on the training and test samples, in more than $90\%$ of the cases. Next, we concentrate on further improving the accuracy of the methods that give the best results by performing some balancing in the feature space. We notice that changes in the  eigenvalues contribute less to the classification results because the sum of all eigenvalues is equal to 2, but each of the angles corresponding to the eigenvectors varies from 0 to 1. To achieve more homogeneity, we use the following transformation:
\begin{equation}\label{fano}
    \lambda'=0.5 (\lambda -0.5) 
    \begin{pmatrix}
1 & 1 & 1 & 1\\
1 & -1 & -1 & 1\\
-1 & 1 & -1 & 1\\
-1 & -1 & 1 & 1
\end{pmatrix}
\end{equation}

\begin{figure*}[h]
  \includegraphics[width=0.7\textwidth]{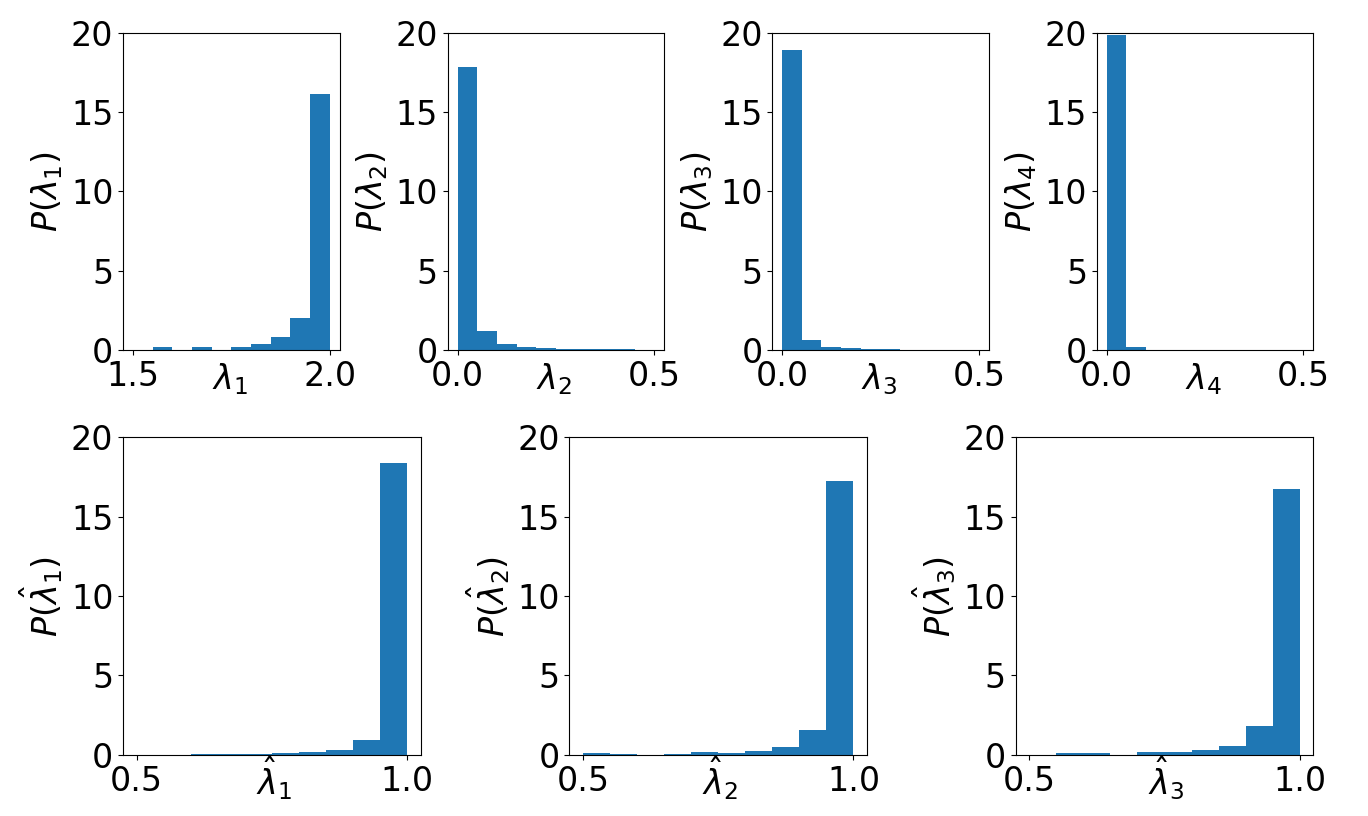}
  \caption{The distribution of the eigenvalues of Choi matrices before (top row) and after (bottom row) the transformation, Eq.~(17).} \label{fig:10}
\end{figure*}

After performing the transformation, we get three non-zero values, which are distributed from 0 to 1. Fig.~\ref{fig:10} shows the distributions of eigenvalues before and after the transformation. 

\begin{figure*}[h]
  \includegraphics[width=0.8\textwidth]{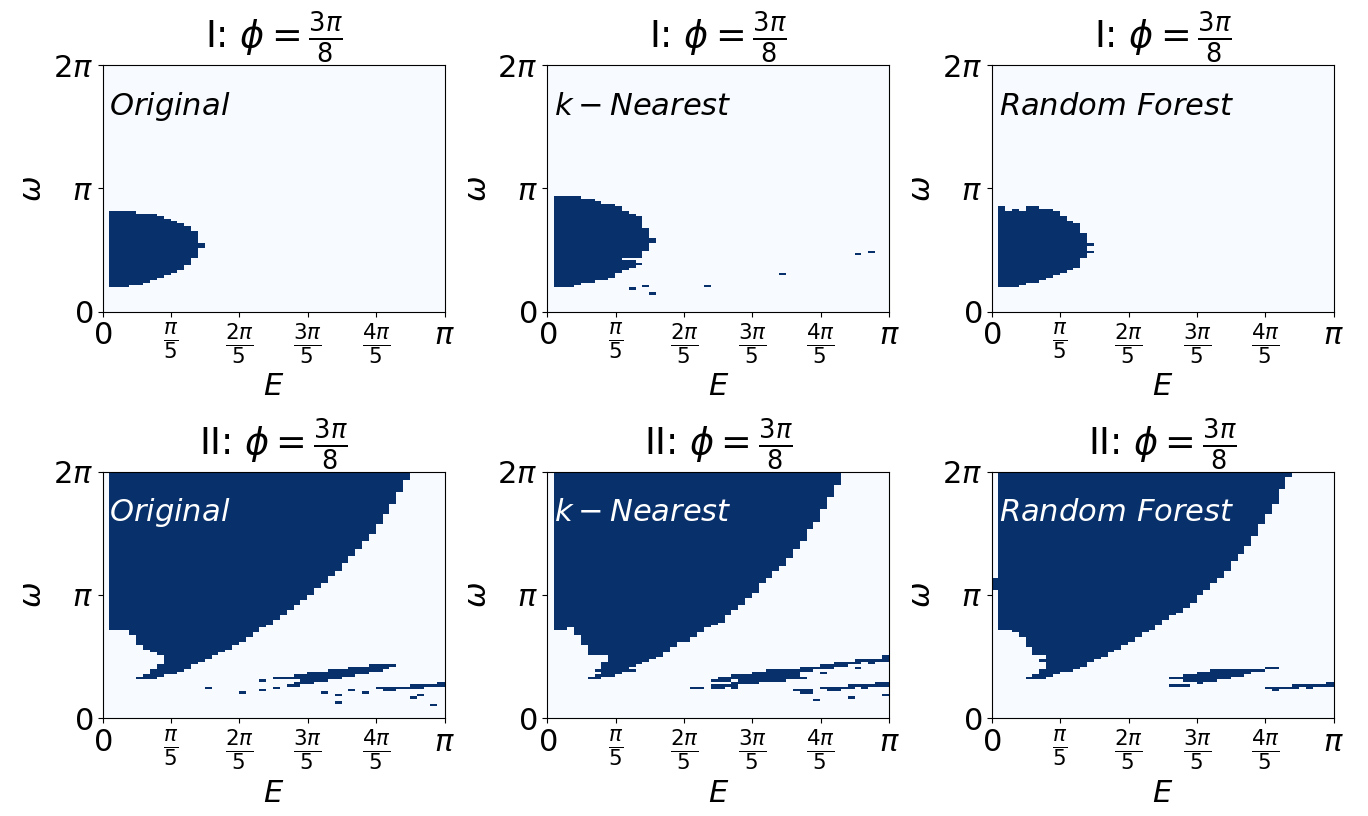}
  \caption{'Yes/no' partition of the parameter space for Problems~I (top row) and II (bottom row) for different values of phase shift. Partitions are obtained by using the test, Eq.~(13) [left] and ML methods, the Nearest neighbors method [center] and the Random forest method [right]. For the ML methods the parametrization based on the eigenvalues, normalized with the transformation (17), and eigenvectors of Choi matrices, is used.} \label{fig:11}
\end{figure*}

The obtained results demonstrate  that after the transformation, the first eigenvalue does no longer dominates over the rest of the eigenvalues. Further on, we scale the values of the angles from the interval $[0, 1]$  to the interval $[0, 0.25]$ in order to increase the importance of the contribution of eigenvalues to the classification results. All this improves the accuracy even of the methods that previously did not work well. The final results are summarized in Table~1 and Fig.~\ref{fig:11}. The diagram shown in Fig.~\ref{fig:11} highlights that the classification errors have local characters and there are no wrong-answer regions appear.

\begin{table*}[]
\begin{ruledtabular}
\begin{tabular}{ccccccc}
Method                        & Train \% & Test \% & Train f1 & Test f1 & Train AUC & Test AUC \\ \hline
Random Forest (Decomposition) & 0.998    & 0.990   & 0.999    & 0.993   & 0.998     & 0.987    \\ 
AdaBoost (Decomposition)      & 0.999    & 0.984   & 0.999    & 0.989   & 0.999     & 0.980    \\ 
AdaBoost (with Normalization) & 0.999    & 0.989   & 0.999    & 0.992   & 0.999     & 0.987    \\ 
\end{tabular}
\caption{\label{tab:1}The best achieved classification accuracy metrics for a dataset of the eigenvalues and eigenvectors of Choi matrices. Samples from Problem~I and Problem~II were used for training and testing.}
\end{ruledtabular}
\end{table*}

\section{Conclusion \label{sec:6}}
In this work, we estimated the potential of Machine Learning (ML) methods as tools to analyze the Floquet-Lindbladian (FL)  problem. We put the emphasis on finding appropriate feature space for which it is possible to construct and train high-accuracy classifiers. As a start, we considered the feature space constructed by using the eigenvalues of the Choi matrices and found that even though good accuracy can be achieved with one model, it is not possible to generalize the obtained classifiers to another problem. Next, we use the whole sets of elements of the Choi matrices. However, this approach did not yield encouraging results either. Finally, by taking into account all the elements of the eigenset, that are eigenvalues and eigenvectors of the Choi matrix, we managed to develop a procedure to purge and normalize data, which allowed us to reach more than $90\%$ of the classification accuracy  when solving the FL problem for both models.


Even though the results we obtained are encouraging and motivating, we should not overestimate the perspectives since we are having a situation rather typical to the ML and AI fields when the results are usually very promising for small-scale test problems. First of all, we cannot prove that the developed schemes would work if the problem set-up is substantially modified. Next, it is not known whether it will be possible to generalize the results to higher values of $N$ (which is our main motivation). 

However, our results provide insight
into the mathematical nature of the original problem~\cite{math}. For example, the fact that, by taking into account not only eigenvalues of the Choi matrix but also eigenvectors, we have obtained substantially better accuracy, tells that some physically relevant information about Markovianity of the map is encoded in the eigenvectors of its Choi matrix.  Since  the test, Eq.~(13), contains matrices derived from the eigenelements of the Floquet map, we could assume that eigenvectors of the Choi matrix also bear information about the degree of Markoviaity of the original Floquet map. Our results validate this assumption and thus constitute, as we believe, a step in the understanding of the Floquet-Lindbladian problem.


\begin{acknowledgments}
This research was funded by the Ministry of Science and Higher Education of the Russian Federation, agreement number 075-15-2020-808.
The numerical experiments
were  performed on the supercomputer “Lomonosov-2” (Moscow State University) and on the supercomputer "Lobachevsky" (Lobachevsky University of Nizhny Novgorod).
\end{acknowledgments}

\section*{Data Availability}
Data sharing is not applicable to this article as no new data were created or analyzed in this study.

\section*{References}
\bibliography{article_ref}

\begin{thebibliography}{38}%
\makeatletter
\providecommand \@ifxundefined [1]{%
 \@ifx{#1\undefined}
}%
\providecommand \@ifnum [1]{%
 \ifnum #1\expandafter \@firstoftwo
 \else \expandafter \@secondoftwo
 \fi
}%
\providecommand \@ifx [1]{%
 \ifx #1\expandafter \@firstoftwo
 \else \expandafter \@secondoftwo
 \fi
}%
\providecommand \natexlab [1]{#1}%
\providecommand \enquote  [1]{``#1''}%
\providecommand \bibnamefont  [1]{#1}%
\providecommand \bibfnamefont [1]{#1}%
\providecommand \citenamefont [1]{#1}%
\providecommand \href@noop [0]{\@secondoftwo}%
\providecommand \href [0]{\begingroup \@sanitize@url \@href}%
\providecommand \@href[1]{\@@startlink{#1}\@@href}%
\providecommand \@@href[1]{\endgroup#1\@@endlink}%
\providecommand \@sanitize@url [0]{\catcode `\\12\catcode `\$12\catcode
  `\&12\catcode `\#12\catcode `\^12\catcode `\_12\catcode `\%12\relax}%
\providecommand \@@startlink[1]{}%
\providecommand \@@endlink[0]{}%
\providecommand \url  [0]{\begingroup\@sanitize@url \@url }%
\providecommand \@url [1]{\endgroup\@href {#1}{\urlprefix }}%
\providecommand \urlprefix  [0]{URL }%
\providecommand \Eprint [0]{\href }%
\providecommand \doibase [0]{http://dx.doi.org/}%
\providecommand \selectlanguage [0]{\@gobble}%
\providecommand \bibinfo  [0]{\@secondoftwo}%
\providecommand \bibfield  [0]{\@secondoftwo}%
\providecommand \translation [1]{[#1]}%
\providecommand \BibitemOpen [0]{}%
\providecommand \bibitemStop [0]{}%
\providecommand \bibitemNoStop [0]{.\EOS\space}%
\providecommand \EOS [0]{\spacefactor3000\relax}%
\providecommand \BibitemShut  [1]{\csname bibitem#1\endcsname}%
\let\auto@bib@innerbib\@empty
\bibitem [{\citenamefont {Elfving}(1937)}]{elfing}%
  \BibitemOpen
  \bibfield  {author} {\bibinfo {author} {\bibfnamefont {G.}~\bibnamefont
  {Elfving}},\ }\bibfield  {title} {\enquote {\bibinfo {title} {{Zur Theorie
  der Markoffschen Ketten}},}\ }\href@noop {} {\bibfield  {journal} {\bibinfo
  {journal} {Acta Societatis Scientiarum Fennicae}\ }\textbf {\bibinfo {volume}
  {2}},\ \bibinfo {pages} {4–11} (\bibinfo {year} {1937})}\BibitemShut
  {NoStop}%
\bibitem [{\citenamefont {Wolf}\ \emph {et~al.}(2008)\citenamefont {Wolf},
  \citenamefont {Eisert}, \citenamefont {Cubitt},\ and\ \citenamefont
  {Cirac}}]{X1}%
  \BibitemOpen
  \bibfield  {author} {\bibinfo {author} {\bibfnamefont {M.}~\bibnamefont
  {Wolf}}, \bibinfo {author} {\bibfnamefont {J.}~\bibnamefont {Eisert}},
  \bibinfo {author} {\bibfnamefont {T.}~\bibnamefont {Cubitt}}, \ and\ \bibinfo
  {author} {\bibfnamefont {J.}~\bibnamefont {Cirac}},\ }\bibfield  {title}
  {\enquote {\bibinfo {title} {{Assessing non-Markovian quantum dynamics}},}\
  }\href {\doibase 10.1103/PhysRevLett.101.150402} {\bibfield  {journal}
  {\bibinfo  {journal} {Physical review letters}\ }\textbf {\bibinfo {volume}
  {101}},\ \bibinfo {pages} {150402} (\bibinfo {year} {2008})}\BibitemShut
  {NoStop}%
\bibitem [{\citenamefont {Cubitt}, \citenamefont {Eisert},\ and\ \citenamefont
  {Wolf}(2012)}]{X2}%
  \BibitemOpen
  \bibfield  {author} {\bibinfo {author} {\bibfnamefont {T.~S.}\ \bibnamefont
  {Cubitt}}, \bibinfo {author} {\bibfnamefont {J.}~\bibnamefont {Eisert}}, \
  and\ \bibinfo {author} {\bibfnamefont {M.~M.}\ \bibnamefont {Wolf}},\
  }\bibfield  {title} {\enquote {\bibinfo {title} {{The complexity of relating
  quantum channels to master equations}},}\ }\href {\doibase
  10.1007/s00220-011-1402-y} {\bibfield  {journal} {\bibinfo  {journal}
  {Communications in Mathematical Physics}\ }\textbf {\bibinfo {volume}
  {310}},\ \bibinfo {pages} {383--418} (\bibinfo {year} {2012})}\BibitemShut
  {NoStop}%
\bibitem [{\citenamefont {Garey}\ and\ \citenamefont {Johnson}(1979)}]{np}%
  \BibitemOpen
  \bibfield  {author} {\bibinfo {author} {\bibfnamefont {M.~R.}\ \bibnamefont
  {Garey}}\ and\ \bibinfo {author} {\bibfnamefont {D.~S.}\ \bibnamefont
  {Johnson}},\ }\href@noop {} {\emph {\bibinfo {title} {{Computers and
  intractability: A guide to the theory of NP-Completeness (Series of Books in
  the Mathematical Sciences)}}}},\ \bibinfo {edition} {first edition}\ ed.\
  (\bibinfo  {publisher} {W. H. Freeman},\ \bibinfo {year} {1979})\BibitemShut
  {NoStop}%
\bibitem [{\citenamefont {Volokitin}\ \emph {et~al.}(2022)\citenamefont
  {Volokitin}, \citenamefont {Kozinov}, \citenamefont {Liniov}, \citenamefont
  {Yusipov}, \citenamefont {Veselov}, \citenamefont {Zolotykh}, \citenamefont
  {Ivanchenko}, \citenamefont {Meyerov},\ and\ \citenamefont
  {Denisov}}]{unpub}%
  \BibitemOpen
  \bibfield  {author} {\bibinfo {author} {\bibfnamefont {V.}~\bibnamefont
  {Volokitin}}, \bibinfo {author} {\bibfnamefont {E.}~\bibnamefont {Kozinov}},
  \bibinfo {author} {\bibfnamefont {A.}~\bibnamefont {Liniov}}, \bibinfo
  {author} {\bibfnamefont {I.}~\bibnamefont {Yusipov}}, \bibinfo {author}
  {\bibfnamefont {S.}~\bibnamefont {Veselov}}, \bibinfo {author} {\bibfnamefont
  {N.}~\bibnamefont {Zolotykh}}, \bibinfo {author} {\bibfnamefont
  {M.}~\bibnamefont {Ivanchenko}}, \bibinfo {author} {\bibfnamefont
  {I.}~\bibnamefont {Meyerov}}, \ and\ \bibinfo {author} {\bibfnamefont
  {S.}~\bibnamefont {Denisov}},\ }\bibfield  {title} {\enquote {\bibinfo
  {title} {{Is there a Lindbladian? Implementation of the test}},}\ }\href@noop
  {} {\bibfield  {journal} {\bibinfo  {journal} {(in preparation)}\ } (\bibinfo
  {year} {2022})}\BibitemShut {NoStop}%
\bibitem [{\citenamefont {Holthaus}(2015)}]{Holthaus2015}%
  \BibitemOpen
  \bibfield  {author} {\bibinfo {author} {\bibfnamefont {M.}~\bibnamefont
  {Holthaus}},\ }\bibfield  {title} {\enquote {\bibinfo {title} {{Floquet
  engineering with quasienergy bands of periodically driven optical
  lattices}},}\ }\href {\doibase 10.1088/0953-4075/49/1/013001} {\bibfield
  {journal} {\bibinfo  {journal} {Journal of Physics B: Atomic, Molecular and
  Optical Physics}\ }\textbf {\bibinfo {volume} {49}},\ \bibinfo {pages}
  {013001} (\bibinfo {year} {2015})}\BibitemShut {NoStop}%
\bibitem [{\citenamefont {Bukov}, \citenamefont {D'Alessio},\ and\
  \citenamefont {Polkovnikov}(2015)}]{Bukov2015}%
  \BibitemOpen
  \bibfield  {author} {\bibinfo {author} {\bibfnamefont {M.}~\bibnamefont
  {Bukov}}, \bibinfo {author} {\bibfnamefont {L.}~\bibnamefont {D'Alessio}}, \
  and\ \bibinfo {author} {\bibfnamefont {A.}~\bibnamefont {Polkovnikov}},\
  }\bibfield  {title} {\enquote {\bibinfo {title} {{Universal high-frequency
  behavior of periodically driven systems: from dynamical stabilization to
  Floquet engineering}},}\ }\href {\doibase 10.1080/00018732.2015.1055918}
  {\bibfield  {journal} {\bibinfo  {journal} {Advances in Physics}\ }\textbf
  {\bibinfo {volume} {64}},\ \bibinfo {pages} {139--226} (\bibinfo {year}
  {2015})}\BibitemShut {NoStop}%
\bibitem [{\citenamefont {Schnell}, \citenamefont {Eckardt},\ and\
  \citenamefont {Denisov}(2020)}]{X3}%
  \BibitemOpen
  \bibfield  {author} {\bibinfo {author} {\bibfnamefont {A.}~\bibnamefont
  {Schnell}}, \bibinfo {author} {\bibfnamefont {A.}~\bibnamefont {Eckardt}}, \
  and\ \bibinfo {author} {\bibfnamefont {S.}~\bibnamefont {Denisov}},\
  }\bibfield  {title} {\enquote {\bibinfo {title} {{Is there a Floquet
  Lindbladian?}}}\ }\href {\doibase 10.1103/PhysRevB.101.100301} {\bibfield
  {journal} {\bibinfo  {journal} {Physical Review B}\ }\textbf {\bibinfo
  {volume} {101}},\ \bibinfo {pages} {100301} (\bibinfo {year}
  {2020})}\BibitemShut {NoStop}%
\bibitem [{\citenamefont {Yusipov}\ \emph {et~al.}(2021)\citenamefont
  {Yusipov}, \citenamefont {Volokitin}, \citenamefont {Liniov}, \citenamefont
  {Ivanchenko}, \citenamefont {Meyerov},\ and\ \citenamefont {Denisov}}]{X4}%
  \BibitemOpen
  \bibfield  {author} {\bibinfo {author} {\bibfnamefont {I.~I.}\ \bibnamefont
  {Yusipov}}, \bibinfo {author} {\bibfnamefont {V.~D.}\ \bibnamefont
  {Volokitin}}, \bibinfo {author} {\bibfnamefont {A.~V.}\ \bibnamefont
  {Liniov}}, \bibinfo {author} {\bibfnamefont {M.~V.}\ \bibnamefont
  {Ivanchenko}}, \bibinfo {author} {\bibfnamefont {I.~B.}\ \bibnamefont
  {Meyerov}}, \ and\ \bibinfo {author} {\bibfnamefont {S.~V.}\ \bibnamefont
  {Denisov}},\ }\bibfield  {title} {\enquote {\bibinfo {title} {{Machine
  Learning versus Semidefinite Programming approach to a particular problem of
  the theory of open quantum systems}},}\ }\href {\doibase
  10.1134/S199508022107026X} {\bibfield  {journal} {\bibinfo  {journal}
  {Lobachevskii Journal of Mathematics}\ }\textbf {\bibinfo {volume} {42}},\
  \bibinfo {pages} {1622--1629} (\bibinfo {year} {2021})}\BibitemShut {NoStop}%
\bibitem [{\citenamefont {Zhang}(2016)}]{X5}%
  \BibitemOpen
  \bibfield  {author} {\bibinfo {author} {\bibfnamefont {Z.}~\bibnamefont
  {Zhang}},\ }\bibfield  {title} {\enquote {\bibinfo {title} {{Introduction to
  Machine Learning: K-Nearest Neighbors}},}\ }\href {\doibase
  10.21037/atm.2016.03.37} {\bibfield  {journal} {\bibinfo  {journal} {Annals
  of Translational Medicine}\ }\textbf {\bibinfo {volume} {4}},\ \bibinfo
  {pages} {218--218} (\bibinfo {year} {2016})}\BibitemShut {NoStop}%
\bibitem [{\citenamefont {Kramer}(2013)}]{X6}%
  \BibitemOpen
  \bibfield  {author} {\bibinfo {author} {\bibfnamefont {O.}~\bibnamefont
  {Kramer}},\ }\enquote {\bibinfo {title} {{K-Nearest Neighbors}},}\ in\ \href
  {\doibase 10.1007/978-3-642-38652-7_2} {\emph {\bibinfo {booktitle}
  {Dimensionality reduction with unsupervised Nearest Neighbors}}}\ (\bibinfo
  {publisher} {Springer Berlin Heidelberg},\ \bibinfo {address} {Berlin,
  Heidelberg},\ \bibinfo {year} {2013})\ pp.\ \bibinfo {pages}
  {13--23}\BibitemShut {NoStop}%
\bibitem [{\citenamefont {Suykens}\ and\ \citenamefont
  {Vandewalle}(1999)}]{X7}%
  \BibitemOpen
  \bibfield  {author} {\bibinfo {author} {\bibfnamefont {J.~A.~K.}\
  \bibnamefont {Suykens}}\ and\ \bibinfo {author} {\bibfnamefont
  {J.}~\bibnamefont {Vandewalle}},\ }\bibfield  {title} {\enquote {\bibinfo
  {title} {{Least Squares Support Vector Machine Classifiers}},}\ }\href
  {\doibase 10.1023/A:1018628609742} {\bibfield  {journal} {\bibinfo  {journal}
  {Neural Processing Letters}\ }\textbf {\bibinfo {volume} {9}},\ \bibinfo
  {pages} {293--300} (\bibinfo {year} {1999})}\BibitemShut {NoStop}%
\bibitem [{\citenamefont {Amari}\ and\ \citenamefont {Wu}(2001)}]{X8}%
  \BibitemOpen
  \bibfield  {author} {\bibinfo {author} {\bibfnamefont {S.}~\bibnamefont
  {Amari}}\ and\ \bibinfo {author} {\bibfnamefont {S.}~\bibnamefont {Wu}},\
  }\bibfield  {title} {\enquote {\bibinfo {title} {{Improving support Vector
  Machine Classifiers by modifying kernel functions}},}\ }\href {\doibase
  10.1016/S0893-6080(99)00032-5} {\bibfield  {journal} {\bibinfo  {journal}
  {Neural Networks}\ }\textbf {\bibinfo {volume} {12}},\ \bibinfo {pages}
  {783--789} (\bibinfo {year} {2001})}\BibitemShut {NoStop}%
\bibitem [{\citenamefont {Dagher}(2008)}]{X9}%
  \BibitemOpen
  \bibfield  {author} {\bibinfo {author} {\bibfnamefont {I.}~\bibnamefont
  {Dagher}},\ }\bibfield  {title} {\enquote {\bibinfo {title} {{Quadratic
  kernel-Free non-linear Support Vector Machine}},}\ }\href {\doibase
  10.1007/s10898-007-9162-0} {\bibfield  {journal} {\bibinfo  {journal}
  {Journal of Global Optimization}\ }\textbf {\bibinfo {volume} {41}},\
  \bibinfo {pages} {15--30} (\bibinfo {year} {2008})}\BibitemShut {NoStop}%
\bibitem [{\citenamefont {Breuer}(2004)}]{Breuer2004}%
  \BibitemOpen
  \bibfield  {author} {\bibinfo {author} {\bibfnamefont {H.-P.}\ \bibnamefont
  {Breuer}},\ }\bibfield  {title} {\enquote {\bibinfo {title} {{Genuine quantum
  trajectories for non-Markovian processes}},}\ }\href {\doibase
  10.1103/PhysRevA.70.012106} {\bibfield  {journal} {\bibinfo  {journal}
  {Physical Review A}\ }\textbf {\bibinfo {volume} {70}},\ \bibinfo {pages}
  {012106} (\bibinfo {year} {2004})}\BibitemShut {NoStop}%
\bibitem [{\citenamefont {Breuer}, \citenamefont {Laine},\ and\ \citenamefont
  {Piilo}(2009)}]{BreuerEtAl09}%
  \BibitemOpen
  \bibfield  {author} {\bibinfo {author} {\bibfnamefont {H.-P.}\ \bibnamefont
  {Breuer}}, \bibinfo {author} {\bibfnamefont {E.-M.}\ \bibnamefont {Laine}}, \
  and\ \bibinfo {author} {\bibfnamefont {J.}~\bibnamefont {Piilo}},\ }\bibfield
   {title} {\enquote {\bibinfo {title} {{Measure for the degree of
  non-Markovian behavior of quantum processes in open systems}},}\ }\href
  {\doibase 10.1103/PhysRevLett.103.210401} {\bibfield  {journal} {\bibinfo
  {journal} {Physical Review Letters}\ }\textbf {\bibinfo {volume} {103}},\
  \bibinfo {pages} {210401} (\bibinfo {year} {2009})}\BibitemShut {NoStop}%
\bibitem [{\citenamefont {Gorini}, \citenamefont {Kossakowski},\ and\
  \citenamefont {Sudarshan}(1976)}]{gorini}%
  \BibitemOpen
  \bibfield  {author} {\bibinfo {author} {\bibfnamefont {V.}~\bibnamefont
  {Gorini}}, \bibinfo {author} {\bibfnamefont {A.}~\bibnamefont {Kossakowski}},
  \ and\ \bibinfo {author} {\bibfnamefont {E.~C.~G.}\ \bibnamefont
  {Sudarshan}},\ }\bibfield  {title} {\enquote {\bibinfo {title} {{Completely
  positive dynamical semigroups of N-Level systems}},}\ }\href {\doibase
  10.1063/1.522979} {\bibfield  {journal} {\bibinfo  {journal} {Journal of
  Mathematical Physics}\ }\textbf {\bibinfo {volume} {17}},\ \bibinfo {pages}
  {821} (\bibinfo {year} {1976})}\BibitemShut {NoStop}%
\bibitem [{\citenamefont {Lindblad}(1976)}]{lindblad}%
  \BibitemOpen
  \bibfield  {author} {\bibinfo {author} {\bibfnamefont {G.}~\bibnamefont
  {Lindblad}},\ }\bibfield  {title} {\enquote {\bibinfo {title} {{On the
  generators of quantum dynamical semigroups}},}\ }\href {\doibase
  10.1007/bf01608499} {\bibfield  {journal} {\bibinfo  {journal}
  {Communications of Mathematical Physics}\ }\textbf {\bibinfo {volume} {48}},\
  \bibinfo {pages} {119--130} (\bibinfo {year} {1976})}\BibitemShut {NoStop}%
\bibitem [{\citenamefont {Khachiyan}\ and\ \citenamefont
  {Porkolab}(1997)}]{Ellipsoid}%
  \BibitemOpen
  \bibfield  {author} {\bibinfo {author} {\bibfnamefont {L.}~\bibnamefont
  {Khachiyan}}\ and\ \bibinfo {author} {\bibfnamefont {L.}~\bibnamefont
  {Porkolab}},\ }\bibfield  {title} {\enquote {\bibinfo {title} {{Computing
  integral points in Convex semi-algebraic sets}},}\ }in\ \href {\doibase
  10.1109/SFCS.1997.646105} {\emph {\bibinfo {booktitle} {Proceedings 38th
  Annual Symposium on Foundations of Computer Science}}}\ (\bibinfo {year}
  {1997})\ pp.\ \bibinfo {pages} {162--171}\BibitemShut {NoStop}%
\bibitem [{\citenamefont {Flum}\ and\ \citenamefont {Grohe}(2006)}]{parameter}%
  \BibitemOpen
  \bibfield  {author} {\bibinfo {author} {\bibfnamefont {J.}~\bibnamefont
  {Flum}}\ and\ \bibinfo {author} {\bibfnamefont {M.}~\bibnamefont {Grohe}},\
  }\href@noop {} {\emph {\bibinfo {title} {{Parameterized Complexity Theory
  (Texts in Theoretical Computer Science. An EATCS Series)}}}}\ (\bibinfo
  {publisher} {Springer-Verlag},\ \bibinfo {address} {Berlin, Heidelberg},\
  \bibinfo {year} {2006})\BibitemShut {NoStop}%
\bibitem [{\citenamefont {Carleo}\ \emph {et~al.}(2019)\citenamefont {Carleo},
  \citenamefont {Cirac}, \citenamefont {Cranmer}, \citenamefont {Daudet},
  \citenamefont {Schuld}, \citenamefont {Tishby}, \citenamefont
  {Vogt-Maranto},\ and\ \citenamefont {Zdeborov\'a}}]{rmp2019}%
  \BibitemOpen
  \bibfield  {author} {\bibinfo {author} {\bibfnamefont {G.}~\bibnamefont
  {Carleo}}, \bibinfo {author} {\bibfnamefont {I.}~\bibnamefont {Cirac}},
  \bibinfo {author} {\bibfnamefont {K.}~\bibnamefont {Cranmer}}, \bibinfo
  {author} {\bibfnamefont {L.}~\bibnamefont {Daudet}}, \bibinfo {author}
  {\bibfnamefont {M.}~\bibnamefont {Schuld}}, \bibinfo {author} {\bibfnamefont
  {N.}~\bibnamefont {Tishby}}, \bibinfo {author} {\bibfnamefont
  {L.}~\bibnamefont {Vogt-Maranto}}, \ and\ \bibinfo {author} {\bibfnamefont
  {L.}~\bibnamefont {Zdeborov\'a}},\ }\bibfield  {title} {\enquote {\bibinfo
  {title} {{Machine Learning and the Physical Sciences}},}\ }\href {\doibase
  10.1103/RevModPhys.91.045002} {\bibfield  {journal} {\bibinfo  {journal}
  {Reviews of Modern Physics}\ }\textbf {\bibinfo {volume} {91}},\ \bibinfo
  {pages} {045002} (\bibinfo {year} {2019})}\BibitemShut {NoStop}%
\bibitem [{\citenamefont {Choi}(1975)}]{choi}%
  \BibitemOpen
  \bibfield  {author} {\bibinfo {author} {\bibfnamefont {M.-D.}\ \bibnamefont
  {Choi}},\ }\bibfield  {title} {\enquote {\bibinfo {title} {{Completely
  positive linear maps on complex matrices}},}\ }\href {\doibase
  https://doi.org/10.1016/0024-3795(75)90075-0} {\bibfield  {journal} {\bibinfo
   {journal} {Linear Algebra and its Applications}\ }\textbf {\bibinfo {volume}
  {10}},\ \bibinfo {pages} {285--290} (\bibinfo {year} {1975})}\BibitemShut
  {NoStop}%
\bibitem [{\citenamefont {{\.Z}yczkowski}\ and\ \citenamefont
  {Bengtsson}(2004)}]{duala}%
  \BibitemOpen
  \bibfield  {author} {\bibinfo {author} {\bibfnamefont {K.}~\bibnamefont
  {{\.Z}yczkowski}}\ and\ \bibinfo {author} {\bibfnamefont {I.}~\bibnamefont
  {Bengtsson}},\ }\bibfield  {title} {\enquote {\bibinfo {title} {{On duality
  between quantum states and quantum maps}},}\ }\href@noop {} {\bibfield
  {journal} {\bibinfo  {journal} {Open Systems \& Information Dynamics}\
  }\textbf {\bibinfo {volume} {11}},\ \bibinfo {pages} {3--42} (\bibinfo {year}
  {2004})}\BibitemShut {NoStop}%
\bibitem [{\citenamefont {{Boyd}}\ \emph {et~al.}(1994)\citenamefont {{Boyd}},
  \citenamefont {{El Ghaoui}}, \citenamefont {{Feron}},\ and\ \citenamefont
  {{Balakrishnan}}}]{control}%
  \BibitemOpen
  \bibfield  {author} {\bibinfo {author} {\bibfnamefont {S.}~\bibnamefont
  {{Boyd}}}, \bibinfo {author} {\bibfnamefont {L.}~\bibnamefont {{El Ghaoui}}},
  \bibinfo {author} {\bibfnamefont {E.}~\bibnamefont {{Feron}}}, \ and\
  \bibinfo {author} {\bibfnamefont {V.}~\bibnamefont {{Balakrishnan}}},\
  }\href@noop {} {\emph {\bibinfo {title} {{Linear matrix inequalities in
  system and control theory}}}}\ (\bibinfo  {publisher} {Society for Industrial
  and Applied Mathematics},\ \bibinfo {year} {1994})\BibitemShut {NoStop}%
\bibitem [{\citenamefont {Ramana}\ and\ \citenamefont
  {Goldman}(1995)}]{spectra}%
  \BibitemOpen
  \bibfield  {author} {\bibinfo {author} {\bibfnamefont {M.}~\bibnamefont
  {Ramana}}\ and\ \bibinfo {author} {\bibfnamefont {A.~J.}\ \bibnamefont
  {Goldman}},\ }\bibfield  {title} {\enquote {\bibinfo {title} {{Some geometric
  results in semidefinite programming}},}\ }\href {\doibase 10.1007/BF01100204}
  {\bibfield  {journal} {\bibinfo  {journal} {Journal of Global Optimization}\
  }\textbf {\bibinfo {volume} {7}},\ \bibinfo {pages} {33--50} (\bibinfo {year}
  {1995})}\BibitemShut {NoStop}%
\bibitem [{\citenamefont {Pedregosa}\ \emph {et~al.}(2011)\citenamefont
  {Pedregosa}, \citenamefont {Varoquaux}, \citenamefont {Gramfort},
  \citenamefont {Michel}, \citenamefont {Thirion}, \citenamefont {Grisel},
  \citenamefont {Blondel}, \citenamefont {Prettenhofer}, \citenamefont {Weiss},
  \citenamefont {Dubourg}, \citenamefont {Vanderplas}, \citenamefont {Passos},
  \citenamefont {Cournapeau}, \citenamefont {Brucher}, \citenamefont {Perrot},\
  and\ \citenamefont {Duchesnay}}]{scikit-learn}%
  \BibitemOpen
  \bibfield  {author} {\bibinfo {author} {\bibfnamefont {F.}~\bibnamefont
  {Pedregosa}}, \bibinfo {author} {\bibfnamefont {G.}~\bibnamefont
  {Varoquaux}}, \bibinfo {author} {\bibfnamefont {A.}~\bibnamefont {Gramfort}},
  \bibinfo {author} {\bibfnamefont {V.}~\bibnamefont {Michel}}, \bibinfo
  {author} {\bibfnamefont {B.}~\bibnamefont {Thirion}}, \bibinfo {author}
  {\bibfnamefont {O.}~\bibnamefont {Grisel}}, \bibinfo {author} {\bibfnamefont
  {M.}~\bibnamefont {Blondel}}, \bibinfo {author} {\bibfnamefont
  {P.}~\bibnamefont {Prettenhofer}}, \bibinfo {author} {\bibfnamefont
  {R.}~\bibnamefont {Weiss}}, \bibinfo {author} {\bibfnamefont
  {V.}~\bibnamefont {Dubourg}}, \bibinfo {author} {\bibfnamefont
  {J.}~\bibnamefont {Vanderplas}}, \bibinfo {author} {\bibfnamefont
  {A.}~\bibnamefont {Passos}}, \bibinfo {author} {\bibfnamefont
  {D.}~\bibnamefont {Cournapeau}}, \bibinfo {author} {\bibfnamefont
  {M.}~\bibnamefont {Brucher}}, \bibinfo {author} {\bibfnamefont
  {M.}~\bibnamefont {Perrot}}, \ and\ \bibinfo {author} {\bibfnamefont
  {E.}~\bibnamefont {Duchesnay}},\ }\bibfield  {title} {\enquote {\bibinfo
  {title} {{Scikit-learn: Machine Learning in Python}},}\ }\href@noop {}
  {\bibfield  {journal} {\bibinfo  {journal} {Journal of Machine Learning
  Research}\ }\textbf {\bibinfo {volume} {12}},\ \bibinfo {pages} {2825--2830}
  (\bibinfo {year} {2011})}\BibitemShut {NoStop}%
\bibitem [{\citenamefont {Safavian}\ and\ \citenamefont
  {Landgrebe}(1991)}]{X10}%
  \BibitemOpen
  \bibfield  {author} {\bibinfo {author} {\bibfnamefont {S.}~\bibnamefont
  {Safavian}}\ and\ \bibinfo {author} {\bibfnamefont {D.}~\bibnamefont
  {Landgrebe}},\ }\bibfield  {title} {\enquote {\bibinfo {title} {{A Survey of
  decision Tree Classifier methodology}},}\ }\href {\doibase 10.1109/21.97458}
  {\bibfield  {journal} {\bibinfo  {journal} {IEEE Transactions on Systems,
  Man, and Cybernetics}\ }\textbf {\bibinfo {volume} {21}},\ \bibinfo {pages}
  {660--674} (\bibinfo {year} {1991})}\BibitemShut {NoStop}%
\bibitem [{\citenamefont {Breiman}\ \emph {et~al.}(1984)\citenamefont
  {Breiman}, \citenamefont {Friedman}, \citenamefont {Olshen},\ and\
  \citenamefont {Stone}}]{X11}%
  \BibitemOpen
  \bibfield  {author} {\bibinfo {author} {\bibfnamefont {L.}~\bibnamefont
  {Breiman}}, \bibinfo {author} {\bibfnamefont {J.~H.}\ \bibnamefont
  {Friedman}}, \bibinfo {author} {\bibfnamefont {R.~A.}\ \bibnamefont
  {Olshen}}, \ and\ \bibinfo {author} {\bibfnamefont {C.~J.}\ \bibnamefont
  {Stone}},\ }\href@noop {} {\emph {\bibinfo {title} {{Classification and
  Regression Trees}}}}\ (\bibinfo  {publisher} {Wadsworth International
  Group},\ \bibinfo {address} {Belmont, CA},\ \bibinfo {year}
  {1984})\BibitemShut {NoStop}%
\bibitem [{\citenamefont {Cutler}, \citenamefont {Cutler},\ and\ \citenamefont
  {Stevens}(2012)}]{X12}%
  \BibitemOpen
  \bibfield  {author} {\bibinfo {author} {\bibfnamefont {A.}~\bibnamefont
  {Cutler}}, \bibinfo {author} {\bibfnamefont {D.~R.}\ \bibnamefont {Cutler}},
  \ and\ \bibinfo {author} {\bibfnamefont {J.~R.}\ \bibnamefont {Stevens}},\
  }\enquote {\bibinfo {title} {{Random Forests}},}\ in\ \href {\doibase
  10.1007/978-1-4419-9326-7_5} {\emph {\bibinfo {booktitle} {Ensemble Machine
  Learning: Methods and Applications}}},\ \bibinfo {editor} {edited by\
  \bibinfo {editor} {\bibfnamefont {C.}~\bibnamefont {Zhang}}\ and\ \bibinfo
  {editor} {\bibfnamefont {Y.}~\bibnamefont {Ma}}}\ (\bibinfo  {publisher}
  {Springer US},\ \bibinfo {address} {Boston, MA},\ \bibinfo {year} {2012})\
  pp.\ \bibinfo {pages} {157--175}\BibitemShut {NoStop}%
\bibitem [{\citenamefont {Breiman}(2001)}]{X13}%
  \BibitemOpen
  \bibfield  {author} {\bibinfo {author} {\bibfnamefont {L.}~\bibnamefont
  {Breiman}},\ }\bibfield  {title} {\enquote {\bibinfo {title} {{Random
  Forests}},}\ }\href {\doibase 10.1023/A:1010933404324} {\bibfield  {journal}
  {\bibinfo  {journal} {Machine Learning}\ }\textbf {\bibinfo {volume} {45}},\
  \bibinfo {pages} {5--32} (\bibinfo {year} {2001})}\BibitemShut {NoStop}%
\bibitem [{\citenamefont {Sanger}(1989)}]{X14}%
  \BibitemOpen
  \bibfield  {author} {\bibinfo {author} {\bibfnamefont {T.~D.}\ \bibnamefont
  {Sanger}},\ }\bibfield  {title} {\enquote {\bibinfo {title} {{Optimal
  Unsupervised Learning in a single-layer linear Feedforward Neural
  Network}},}\ }\href@noop {} {\bibfield  {journal} {\bibinfo  {journal}
  {Neural networks}\ }\textbf {\bibinfo {volume} {2}},\ \bibinfo {pages}
  {459--473} (\bibinfo {year} {1989})}\BibitemShut {NoStop}%
\bibitem [{\citenamefont {Glorot}\ and\ \citenamefont {Bengio}(2010)}]{X15}%
  \BibitemOpen
  \bibfield  {author} {\bibinfo {author} {\bibfnamefont {X.}~\bibnamefont
  {Glorot}}\ and\ \bibinfo {author} {\bibfnamefont {Y.}~\bibnamefont
  {Bengio}},\ }\bibfield  {title} {\enquote {\bibinfo {title} {{Understanding
  the difficulty of training Deep Feedforward Neural Networks}},}\ }in\
  \href@noop {} {\emph {\bibinfo {booktitle} {Proceedings of the Thirteenth
  International Conference on Artificial Intelligence and Statistics}}},\
  \bibinfo {series} {JMLR Proceedings}, Vol.~\bibinfo {volume} {9},\ \bibinfo
  {editor} {edited by\ \bibinfo {editor} {\bibfnamefont {Y.~W.}\ \bibnamefont
  {Teh}}\ and\ \bibinfo {editor} {\bibfnamefont {D.~M.}\ \bibnamefont
  {Titterington}}}\ (\bibinfo  {publisher} {JMLR.org},\ \bibinfo {year}
  {2010})\ pp.\ \bibinfo {pages} {249--256}\BibitemShut {NoStop}%
\bibitem [{\citenamefont {Freund}\ and\ \citenamefont {Schapire}(1997)}]{X17}%
  \BibitemOpen
  \bibfield  {author} {\bibinfo {author} {\bibfnamefont {Y.}~\bibnamefont
  {Freund}}\ and\ \bibinfo {author} {\bibfnamefont {R.~E.}\ \bibnamefont
  {Schapire}},\ }\bibfield  {title} {\enquote {\bibinfo {title} {{A
  decision-theoretic generalization of on-line learning and an application to
  boosting}},}\ }\href {\doibase https://doi.org/10.1006/jcss.1997.1504}
  {\bibfield  {journal} {\bibinfo  {journal} {Journal of Computer and System
  Sciences}\ }\textbf {\bibinfo {volume} {55}},\ \bibinfo {pages} {119--139}
  (\bibinfo {year} {1997})}\BibitemShut {NoStop}%
\bibitem [{\citenamefont {Fukunaga}(2013)}]{X18}%
  \BibitemOpen
  \bibfield  {author} {\bibinfo {author} {\bibfnamefont {K.}~\bibnamefont
  {Fukunaga}},\ }\href@noop {} {\emph {\bibinfo {title} {{Introduction to
  Statistical Pattern Recognition}}}}\ (\bibinfo  {publisher} {Elsevier},\
  \bibinfo {year} {2013})\BibitemShut {NoStop}%
\bibitem [{\citenamefont {Domingos}\ and\ \citenamefont {Pazzani}(1997)}]{X19}%
  \BibitemOpen
  \bibfield  {author} {\bibinfo {author} {\bibfnamefont {P.}~\bibnamefont
  {Domingos}}\ and\ \bibinfo {author} {\bibfnamefont {M.}~\bibnamefont
  {Pazzani}},\ }\bibfield  {title} {\enquote {\bibinfo {title} {{On the
  optimality of the simple Bayesian classifier under zero-one Loss}},}\ }\href
  {\doibase 10.1023/A:1007413511361} {\bibfield  {journal} {\bibinfo  {journal}
  {Machine Learning}\ }\textbf {\bibinfo {volume} {29}},\ \bibinfo {pages}
  {103--130} (\bibinfo {year} {1997})}\BibitemShut {NoStop}%
\bibitem [{\citenamefont {Chicco}\ and\ \citenamefont {Jurman}(2020)}]{X21}%
  \BibitemOpen
  \bibfield  {author} {\bibinfo {author} {\bibfnamefont {D.}~\bibnamefont
  {Chicco}}\ and\ \bibinfo {author} {\bibfnamefont {G.}~\bibnamefont
  {Jurman}},\ }\bibfield  {title} {\enquote {\bibinfo {title} {{The Aavantages
  of the Matthews Correlation Coefficient (MCC) over F1 Score and accuracy in
  Binary Classification Evaluation}},}\ }\href {\doibase
  10.1186/s12864-019-6413-7} {\bibfield  {journal} {\bibinfo  {journal} {BMC
  Genomics}\ }\textbf {\bibinfo {volume} {21}} (\bibinfo {year} {2020}),\
  10.1186/s12864-019-6413-7}\BibitemShut {NoStop}%
\bibitem [{\citenamefont {Fawcett}(2006)}]{X22}%
  \BibitemOpen
  \bibfield  {author} {\bibinfo {author} {\bibfnamefont {T.}~\bibnamefont
  {Fawcett}},\ }\bibfield  {title} {\enquote {\bibinfo {title} {{An
  introduction to ROC analysis}},}\ }\href {\doibase
  https://doi.org/10.1016/j.patrec.2005.10.010} {\bibfield  {journal} {\bibinfo
   {journal} {Pattern Recognition Letters}\ }\textbf {\bibinfo {volume} {27}},\
  \bibinfo {pages} {861--874} (\bibinfo {year} {2006})}\BibitemShut {NoStop}%
\bibitem [{\citenamefont {Davies}\ \emph {et~al.}(2021)\citenamefont {Davies}
  \emph {et~al.}}]{math}%
  \BibitemOpen
  \bibfield  {author} {\bibinfo {author} {\bibfnamefont {A.}~\bibnamefont
  {Davies}} \emph {et~al.},\ }\bibfield  {title} {\enquote {\bibinfo {title}
  {{Advancing mathematics by guiding human intuition with AI}},}\ }\href@noop
  {} {\bibfield  {journal} {\bibinfo  {journal} {Nature}\ }\textbf {\bibinfo
  {volume} {600}},\ \bibinfo {pages} {70--74} (\bibinfo {year}
  {2021})}\BibitemShut {NoStop}%
\end{thebibliography}%

\end{document}